\documentclass{nature}
\usepackage{caption}
\DeclareCaptionType{suppfigure}[Supplementary Figure][List of Supplementary Figures]

\DeclareCaptionType{extendedfigure}[Extended Data Figure][List of Extended Data Figures]

\usepackage{adjustbox}
\usepackage{hyperref}
\usepackage{longtable}
\usepackage{subcaption}
\usepackage{geometry}
\usepackage{threeparttable}
\usepackage{amsmath,amssymb,amsxtra,amsfonts}
\usepackage[utf8]{inputenc} % useful to type directly diacritic characters
\usepackage[T1]{fontenc}    % Use vector fonts, so it zooms properly in on-screen viewing software
\usepackage{color}
\usepackage[most]{tcolorbox}
\hypersetup{
    colorlinks=True,
    linkcolor={blue!100!black},
    citecolor={blue!100!black},
    urlcolor={blue!100!black}
}

\newcommand{\araa}{Annu. Rev. Astron. Astrophys.}   % Annual Review of Astron and Astrophys
\newcommand{\aj}{Astron. J.}   % Astronomical Journal
\newcommand{\apj}{Astrophys. J.}   % Astrophysical Journal
\newcommand{\apjl}{Astrophys. J. Lett.}   % Astrophysical Journal, Letters
\newcommand{\apjs}{Astrophys. J. Suppl. Ser.}   % Astrophysical Journal, Supplement
\newcommand{\ao}{Appl. Opt.}   % Applied Optics
\newcommand{\aap}{Astron. Astrophys.}   % Astronomy and Astrophysics
\newcommand{\mnras}{Mon. Not. R. Astron. Soc.}   % Monthly Notices of the RAS
\newcommand{\nat}{Nature} % Nature
\newcommand{\physrep}{Phys. Rep.}   % Physics Reports
\newcommand{\pasa}{Publ. Astron. Soc. Aust.}   % Publications of the Astron. Soc. of Australia
\newcommand{\pasp}{Publ. Astron. Soc. Pac.}   % Publications of the Astron. Soc. of the Pacific
\newcommand{\rmxaa}{Rev. Mexicana Astron. Astrofis.}   % Revista Mexicana de Astronomia y Astrofisica
\let\OLDthebibliography\thebibliography
\renewcommand\thebibliography[1]{
  \OLDthebibliography{#1}
  \setlength{\parskip}{0pt}
  \setlength{\itemsep}{0pt plus 0.3ex}
}
\usepackage{lineno}
\usepackage{multirow}
\usepackage{xspace}

\newcommand\farcs{\mbox{$.\!^{\prime\prime}$}}

\newcommand{\Msun}{\ensuremath{M_\odot}\xspace}

\newcommand{\Hb}{\textrm{H}\ensuremath{\beta}\xspace}

\newcommand{\OII}{[\textrm{O}~\textsc{ii}]\xspace}
\newcommand{\OIII}{[\textrm{O}~\textsc{iii}]\xspace}

\newcommand{\SiII}{\textrm{Si}~\textsc{ii}\xspace}
\newcommand{\CII}{\textrm{C}~\textsc{ii}\xspace}

\title{First-star imprints in a metal-poor galaxy overdensity near the end of reionization}

\author{Zihao Li$^{1,2}$ \thanks{E-mail: \href{mailto:zihao.li@nbi.ku.dk}{zihao.li@nbi.ku.dk}
},
Koki Kakiichi$^{1,2}$,
Lise Christensen$^{1,2}$,
Zheng Cai$^{3}$,
Valentina D’Odorico$^{4,5}$,
Jorryt Matthee$^{6}$,
Daichi Kashino$^{7}$,
Rongmon Bordoloi$^{8}$,
Ruari Mackenzie$^{9}$,
Trystyn A. M. Berg$^{10}$,
Irene Vanni$^{11,12}$,
Stefania Salvadori$^{11,12}$,
Alessandra Venditti$^{13,14}$,  % orcid 0000-0003-2237-0777
Shiwu Zhang$^{15}$,
Sarah E.~I.~Bosman$^{16,17}$,
Eduardo Bañados$^{17}$,
Frederick B.~Davies$^{17}$,
Xiaohui Fan$^{18}$,
Hyunsung D. Jun$^{19,20}$,
Xiangyu Jin$^{21}$,
Mingyu Li$^{3}$,
Sofía Rojas-Ruiz$^{22}$,
Feige Wang$^{21}$,
Jinyi Yang$^{21}$,
Siwei Zou$^{23,24}$,
Huanian Zhang$^{25}$,
Yongda Zhu$^{18}$
}

\makeatletter
\let\saved@includegraphics\includegraphics
\AtBeginDocument{\let\includegraphics\saved@includegraphics}
\renewenvironment*{figure}{\@float{figure}}{\end@float}
\makeatother

\begin{document}
\maketitle

\begin{affiliations}
\item{Cosmic Dawn Center (DAWN), Denmark}
\item{Niels Bohr Institute, University of Copenhagen, Jagtvej 128, DK2200 Copenhagen N, Denmark}
\item{Department of Astronomy, Tsinghua University, Beijing 100084, China}
\item{INAF - Osservatorio Astronomico di Trieste, Via G. Tiepolo 11, 34143, Trieste, Italy}
%\item{Scuola Normale Superiore, Piazza dei Cavalieri 7, 56126, Pisa, Italy}
\item{IFPU - Institute for Fundamental Physics of the Universe, via Beirut 2, I-34151 Trieste, Italy}
\item{Institute of Science and Technology Austria (ISTA), Am Campus 1, 3400 Klosterneuburg, Austria}
\item{National Astronomical Observatory of Japan (NAOJ), 2-21-1, Osawa, Mitaka, Tokyo 181-8588, Japan}
\item{Department of Physics and Astronomy, North Carolina State University, Raleigh, NC 27695, USA}
\item{Laboratory of Astrophysics, École Polytechnique Fédérale de Lausanne (EPFL), Observatoire de Sauverny, 1290 Versoix, Switzerland}
\item{Department of Physics and Astronomy, Camosun College, 3100 Foul Bay Rd, Victoria,  B.C., V8P 5J2, Canada}
\item{Dipartimento di Fisica e Astrofisica, Università degli Studi di Firenze, Via G. Sansone 1, I-50019 Sesto Fiorentino, Italy}
\item{INAF/Osservatorio Astrofisico di Arcetri, Largo E. Fermi 5, I-50125 Firenze, Italy}
\item{Department of Astronomy, University of Texas at Austin, 2515 Speedway, Stop C1400, Austin, TX 78712, USA}
\item{Cosmic Frontier Center, The University of Texas at Austin, Austin, TX 78712, USA}
\item{Department of Automation, Tsinghua University, Beijing, China, 100084}
\item{Institute for Theoretical Physics, Heidelberg University, Philosophenweg 12, D–69120, Heidelberg, Germany}
\item{Max-Planck-Institut für Astronomie, Königstuhl 17, D-69117 Heidelberg, Germany}
\item{Steward Observatory, University of Arizona, 933 N Cherry Avenue, Tucson, AZ 85721, USA}
\item{Department of Physics, Northwestern College, 101 7th Street SW, Orange City, IA 51041, USA}
\item{School of Physics, Korea Institute for Advanced Study, 85 Hoegiro, Dongdaemun-gu, Seoul 02455, Republic of Korea}
\item{Department of Astronomy, University of Michigan, 1085 S. University Ave., Ann Arbor, MI 48109, USA}
\item{Department of Physics and Astronomy, University of California, Los Angeles, 430 Portola Plaza, Los Angeles, CA 90095, USA}
\item{Chinese Academy of Sciences South America Center for Astronomy, National Astronomical Observatories, CAS, Beijing 100101, China}
\item{Departamento de Astronom\'ia, Universidad de Chile, Casilla 36-D, Santiago, Chile}
\item{Department of Astronomy, Huazhong University of Science and Technology, Wuhan 430074, China}

\begingroup
\endgroup

\end{affiliations}

%--------------------------------------------------------------------------------------------------------------------------------%
\begin{abstract}
The first generation of stars, known as Population III (Pop III), formed from primordial gas consisting solely of hydrogen and helium and is believed to have emerged only a few hundred million years after the Big Bang. Detecting the chemical enrichment of metal-poor circumgalactic gas offers a promising way to trace the enrichment signature of Pop III stars. Along the sightline to the quasar SDSS J0100+2802, a metal absorber at $z = 5.945$, showing over-abundant carbon and silicon compared to solar, has been reported to be consistent with the enrichment pattern of Pop III stars. With the James Webb Space Telescope, we report the discovery of an unusually metal-poor galaxy overdensity of 17 members (mean metallicity $\approx 3\%$ solar) near this metal absorber, which is $\sim 0.4$ dex more metal-poor than coeval galaxies in similarly overdense environments. This less chemically evolved system may have provided favorable conditions for preserving the absorption signatures of Pop III enrichment. We infer a minimum dark matter halo of $\log(M_{\mathrm{h,min}}/M_{\odot})=10.68^{+0.93}_{-1.72}$, supporting late-time Pop III formation at the outskirts of atomic hydrogen cooling halos. Our findings open a promising observational pathway to identify the chemical imprints of the first stars and constrain the conditions for their formation.
\end{abstract}

\begin{figure*}[!t]
  \includegraphics[width=\linewidth]{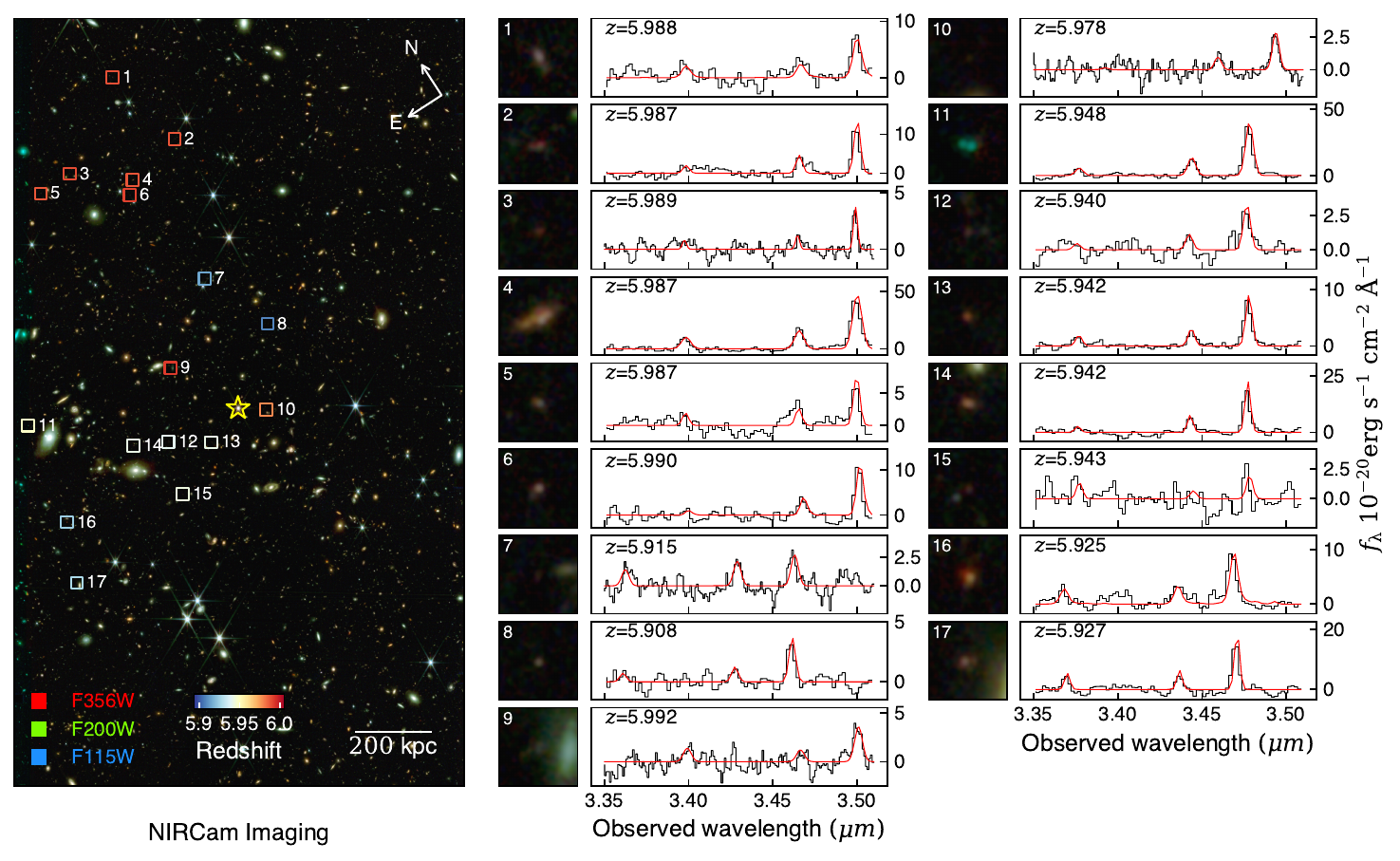}
\caption{\small \textbf{Left panel}: False color image of J0100+2802 field from JWST NIRCam through three bands (F356W, F200W, F115W). The locations of the galaxies in the overdensity are marked with squares, color-coded by redshifts. The central yellow star marks the location of the quasar, along whose sightline the Pop III absorber at $z=5.945$ is intervening. \textbf{Right panels}: One-dimensional grism spectra of each galaxy in the overdensity hosting the Pop III absorber, with zoom-in cutouts shown on the left. The red lines indicate the best-fit Gaussian models to the observed spectra.}\label{fig:J100_fov}
\end{figure*}

The Pop III stars are theoretically predicted to emerge during the cosmic dawn, with typical estimates spanning $z \approx 20-30$, depending on the model\cite{Klessen_23}. Although still model-dependent\cite{Liu_20} and not fully observationally constrained\cite{Fujimoto_25b}, the peak of Pop III star formation is predicted to occur at $z \gtrsim 10$, while direct detection by JWST at $z > 10$ is predicted to be challenging\cite{Schauer_20}, and may be more feasible under extreme conditions, such as in highly overdense regions\cite{Jeon_26}. Formed from metal-free gas, those stars lacked efficient cooling via metals and are therefore theoretically expected to have an initial mass function (IMF) distinct from that of metal-enriched stellar populations\cite{Hirano_14,Chon_24}.
They are predicted to have played an important role in the onset of cosmic reionization through their intense ultraviolet radiation, which ionized the surrounding intergalactic medium (IGM)\cite{Hartwig_22}. 
Depending on their progenitor masses, they ended their lives as supernovae, such as core-collapse supernovae (CCSNe) and pair-instability supernovae (PISNe), enriching the interstellar medium (ISM) and beyond, thereby seeding the early universe with the first heavy elements.

Direct detection of Pop III stars requires identifying a metal-free star-forming environment and unambiguous signatures produced by the stars themselves. 
Tentative He\,{\sc ii}~1640\,\AA\ emission of RX J2129-z8He II at $z \sim 8$ suggests ongoing Pop III star formation producing a hard radiation background in galaxies with stellar masses $M_\star \sim 10^8-10^9~M_\odot$\cite{Wangx_24}.
The strong He\,{\sc ii}~1640\,\AA\ emission observed in Hebe at $z \sim 11$, without associated metal lines\cite{Maiolino_26,Ubler_26}, suggests ongoing Pop III star formation, with Pop III stellar masses of $M_\star \sim (2-60)\times 10^4~\Msun$\cite{Rusta_26}. Recent theoretical work further suggests that Pop III spectral signatures, such as He\,{\sc ii}~1640\,\AA, may still survive in moderately metal-enriched hybrid galaxies\cite{Rusta_25,Venditti_26}, where Pop III stars co-exist with Pop II stars making the extremely metal-poor Pop III candidates particularly interesting\cite{vanzella23, Fujimoto_25,Morishita_25,Fujimoto_25b,Nakajima_25, Vanzella_26}.

\begin{figure*}[!t]
    \centering
    \begin{subfigure}{0.48\textwidth}
        \includegraphics[width=\textwidth]{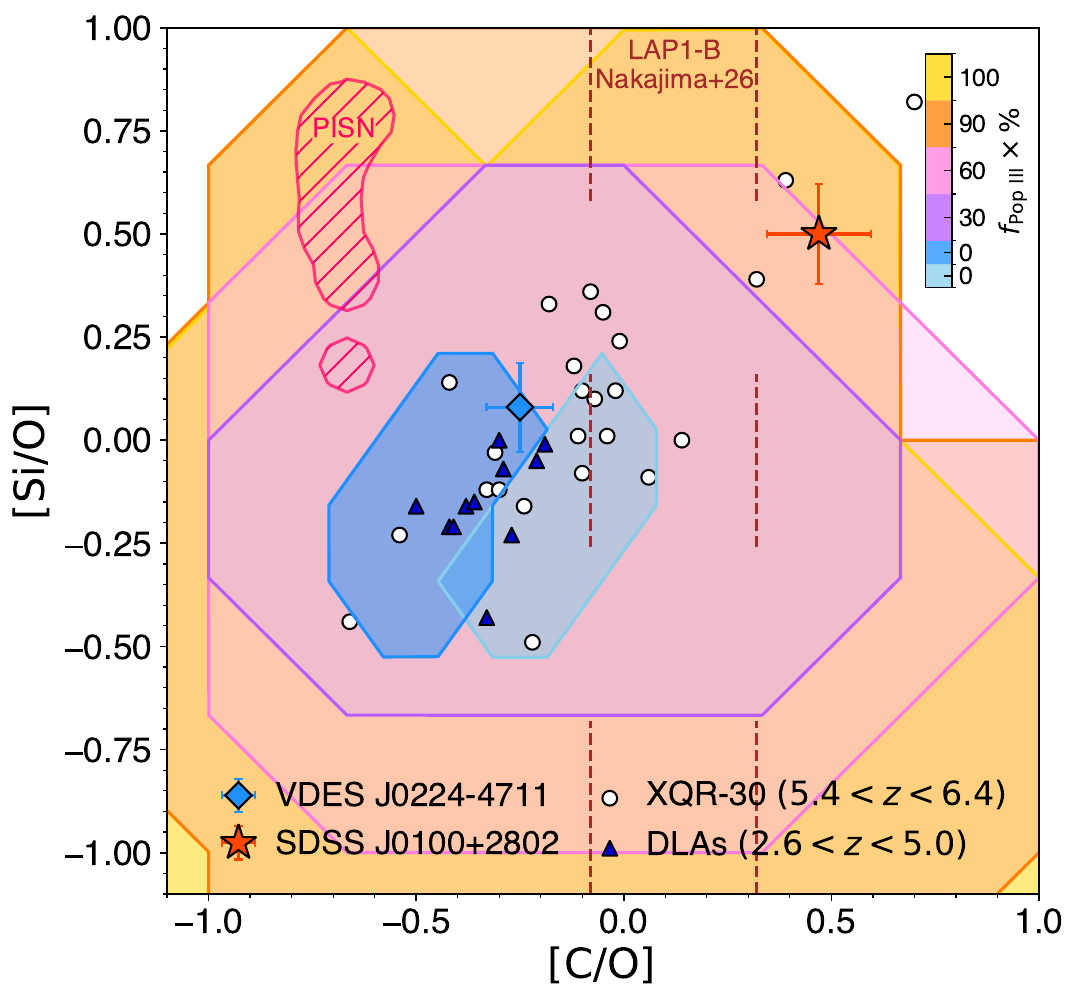}
        \caption{}
        \label{fig:SIO}
    \end{subfigure}
    \begin{subfigure}{0.48\textwidth}
        \includegraphics[width=\textwidth]{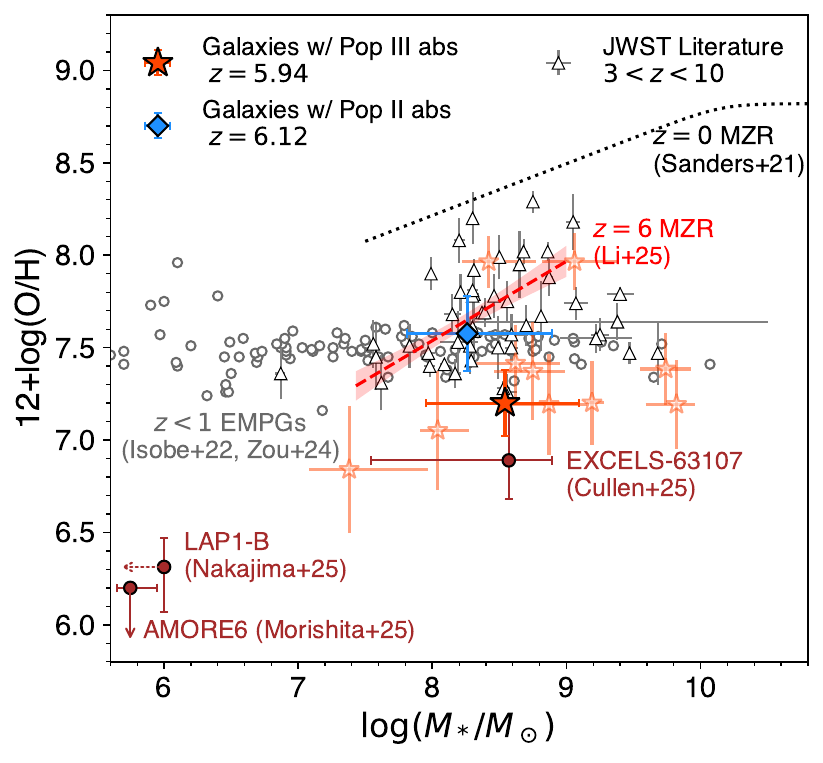}
        \caption{}
        \label{fig:MZR}
    \end{subfigure}
    \caption{\small \textbf{(a)}: Comparison between abundance ratios of [C/O] and [Si/O] in observed absorbers and Pop III mixture model\cite{Vanni_24}. The shaded areas are colored according to the fraction of Pop III CCSNe contributing to the chemical enrichment ($f_{Pop III}$) in the mixture model. The dark blue area represents pure Pop II enrichment based on ref.~\citen{Limongi_18}, while the light blue area shows pure Pop II enrichment from ref.~\citen{Woosley_95}, in both cases $f_{Pop III}=0$. The red hatched contours represent the Pop III PISN model\cite{Heger_02}.
    The open circles represent absorbers from the XQR-30 catalog\cite{Davies_23}, and the open triangles represent metal-poor damped Lyman-$\alpha$ systems (DLAs) at lower redshifts\cite{Welsh_19}. The brown dashed lines indicate the $1\sigma$ interval of [C/O] abundance of an extremely metal-poor galaxy LAP1-B reported by ref.~\citen{Nakajima_25}. \textbf{(b)}: The mass-metallicity relation. The median stack of overdensity member galaxies is shown in red stars, with empty stars representing measurements for individual sources with $\text{SNR}_{\Hb}>3$. We also show the stacked result for the galaxy overdensity in the VDESJ0224-4711 field (blue diamond), which is associated with a Pop II absorber. Literature JWST observations are shown in empty black triangles\cite{Chakraborty_25,Pollock_25}, and extremely metal-poor galaxies EXCELS-63107\cite{Cullen_25}, LAP1-B\cite{Nakajima_25}, and AMORE6\cite{Morishita_25} are highlighted in brown color. We note that LAP1-B has a reported upper limit on its stellar mass of $\sim2700$ \Msun\cite{Nakajima_25}, while we shifted its $x$-axis position upward in this plot for visual clarity.
    The low-redshift extremely metal-poor galaxies (EMPGs)\cite{Isobe_22,Zouh_24} are shown in gray circles for comparison.}
    \label{fig:twoside}
\end{figure*}
Indirect Pop III detection can be achieved through the observation of their nucleosynthetic products as metal absorbers which were dispersed in the environment by stellar winds, supernova explosions and on larger radii, by galactic-scale outflows.
Stellar nucleosynthetic yields provide diagnostic abundance signatures of Pop III\ enrichment. High nitrogen-to-oxygen (N/O) ratios\cite{Ji_24} may indicate pollution by very massive Pop III\ stars ($M_\star > 1000\,M_\odot$)\cite{Nandal_25}, whereas elevated carbon-to-oxygen (C/O) ratios are characteristic of yields from lower-mass Pop III\ stars that end their lives as low-to-normal CCSNe\cite{Ishigaki_14, Jeon_21, Vanni_23, Saccardi_23} with metal mixing and fallback during the explosion. 
The chemical signatures of these Pop III SNe are possibly preserved in long-lived, metal-poor stars in the Milky Way and in ultra-faint dwarf galaxies\cite{Koutsouridou_23}, or in the form of metal absorption features due to the high-redshift circumgalactic medium (CGM) and IGM.
Distant quasars act as bright backlights for detecting those metal absorbers\cite{wolfe05}. Deep JWST and ground-based spectroscopy of high-redshift quasars at $z>6$ are revealing a growing census of metal-enriched systems\cite{Becker_12,christensen23,dodorico23,Davies_23}, and extremely metal-poor absorbers\cite{durovcikova_25}, potentially linked to Pop III star formation. However, conclusive detections of Pop III stars and the environments of their origin remain elusive.

\begin{table*}[h]
\centering
\footnotesize
\begin{threeparttable}
\caption{
\textbf{The measured properties of the two metal absorbers.}
}
\label{tab:ab}
\begin{tabular}{cccccccccc}
\hline\hline
\multirow{2}{*}{Quasar} & \multirow{2}{*}{$z_\text{abs}$} & $\log N_{\text{O}\textsc{i}}$
& $\log N_{\text{C}\textsc{ii}}$ & $\log N_{\text{Si}\textsc{ii}}$
& $\log N_{\text{C}\textsc{iv}}$ & $\log N_{\text{Si}\textsc{iv}}$
& \multirow{2}{*}{[C/O]} & \multirow{2}{*}{[Si/O]} & \multirow{2}{*}{Type$^{a}$} \\
 &  & $(\rm cm^{-2})$ & $(\rm cm^{-2})$ & $(\rm cm^{-2})$ & $(\rm cm^{-2})$ & $(\rm cm^{-2})$ &   &   &   \\
\hline
J0100+2802 & 5.945 & $13.39\pm0.07$ & $13.63\pm0.08$ &
$12.71\pm0.09$ & $<11.8^{b}$ & $<11.4^{b}$ &
$0.47\pm0.12$ & $0.50\pm0.12$ & Pop III \\
J0224-4711 & 6.122 & $14.45\pm0.05$ & $13.39\pm0.09$ &
$13.35\pm0.06$ & $<13.15^{c}$ & -- &
$-0.25\pm0.11$ & $0.08\pm0.08$ & Pop II \\
\hline
\end{tabular}
$^{a}$ The type is determined by consistency with Pop III\ or
Pop II\ enrichment models\cite{Vanni_24}.
$^{b}$ $3\sigma$ upper limit measured at the same velocity as
O\,\textsc{i}.
$^{c}$ Value from ref.~\citen{Zou_24}.
\end{threeparttable}
\end{table*}

JWST slitless spectroscopy is a highly efficient way to detect absorber galaxy environments extending to scales of up to 1 pMpc\cite{Bordoloi_24,Zou_24}. Here, we present results from deep JWST/NIRCam slitless spectroscopy of two luminous quasar fields (Fig. 1), complemented by deep, high-resolution VLT/X-shooter spectroscopy. 
We used the column densities of different elements, including oxygen, carbon, and silicon, derived from the public XQR-30 absorber catalog\cite{Davies_23} and determined their abundances relative to solar reference values (Methods).
By comparing with chemical enrichment models\cite{Vanni_23}, we identified two metal absorbers with abundance patterns consistent with enrichment from Pop III and Pop II supernovae, respectively, along the sightlines of the quasars SDSS J0100+2802 and VDES J0224–4711 (Fig. 2a, see also ref. \citen{sodini24}). Both sightlines are covered by JWST/NIRCam grism observations from the EIGER\cite{Kashino_23} and ASPIRE\cite{Zou_24} programs. Their measured properties are summarized in Table \ref{tab:ab}.

The Pop III chemical-enrichment models in Fig. 2a (Methods) indicate that the absorber at $z_\text{abs}=5.945$ along the sightline of the quasar SDSS J0100+2802 exhibits a distinctly Pop III–like signature: the elevated carbon-to-oxygen and silicon-to-oxygen ratios are consistent with an environment predominantly enriched by the chemical products of Pop III SNe, and we denote this as a ``Pop III absorber" for simplicity. A similarly enhanced C/O ratio has also been reported in the chemically primitive galaxy LAP1-B\cite{Nakajima_25}, whose extremely low oxygen abundance places it in the Pop III regime. This agreement provides an observational consistency check for C/O as a Pop III-sensitive abundance diagnostic.
In contrast, the absorber at $z_\text{abs}=6.122$ toward quasar VDES J0224-4711 exhibits a lower ratio, indistinguishable from typical Pop II SN yields\cite {sodini24}. It is therefore classified as a ``Pop II absorber".
The absorber at $z=6.122$ has also been reported to be associated with an overdensity of five member galaxies\cite{Zou_24}. For the $z=5.945$ absorber, an overdensity within 300 pkpc was also reported\cite{Bordoloi_24,Higginson_25}. Here, we further explore the large-scale environment of the $z=5.945$ absorber. We employed the Friends-of-friends (FoF) clustering algorithm\cite{Li_25} (Methods) on the \OIII catalog detected in the EIGER field\cite{Kashino_25} to identify galaxy groups. We found an overdensity of 17 member galaxies (Fig. 1) spanning a projected scale of $\approx1400$ pkpc and a redshift window of $\Delta z=\pm0.05$, corresponding to a velocity $\Delta v=\pm2161 \rm~km~s^{-1}$. We compute the line-of-sight velocity dispersion of this overdensity to be $1253\pm133\rm~km~s^{-1}$ (Methods).
We show the cross-correlation between the absorber and the associated galaxies in Extended Figure 1. The result indicates the minimum host halo mass of the Pop III absorber as $\log(M_{\mathrm{h,min}}/M_{\odot})=10.68^{+0.93}_{-1.72}$, although measurement uncertainties also allow for the possibility of lower-mass halos down to $\log(M_{\mathrm{h}}/M_{\odot})\sim9$. We thus put the lower limit of the observed halo mass that hosts the Pop III-enriched gas as $\log(M_{\mathrm{h,min}}/M_{\odot})>9$.

We infer the gas-phase metallicity using the \OIII/\Hb\ emission line ratio (R3)\cite{Chakraborty_25} (Methods). To estimate the average metallicity of member galaxies in each overdensity, we construct mean stacked spectra, applying $5\sigma$ clipping to remove outlier pixels, and derive the \OIII/\Hb\ line ratios from the resulting composite spectra. The \OIII/\Hb\ line ratio is then converted to gas-phase metallicity using strong line calibration (Extended Data Figure \ref{fig:R3-Z}). We estimate the stellar masses through spectral energy distribution (SED) fitting (Methods). Fig. 2b shows the mass–metallicity relation (MZR), in comparison with literature observations\cite{}. The overdensity associated with the Pop III absorber lies significantly below the typical MZR in overdense environments\cite{Li_25}, while the Pop II absorber toward VDES J0224-4711 is consistent with MZR predictions. The Pop III absorber therefore appears to trace an unusually metal-poor overdensity, with mean metallicities consistent with recently reported metal-poor environments at comparable epochs\cite{Bolamperti_26}.
For comparison, we show an extremely metal-poor galaxy, EXCELS-63107, with a comparable stellar-mass regime reported in the literature\cite{Cullen_25} in Fig. 2b. Its spectrum indicates an unusually hard ionizing source, and a plausible explanation is the formation of Pop III stars within mildly enriched halos.
In our observations, the metal-deficient nature of the overdensity points to a relatively chemically primitive environment in which early enrichment signatures could be more readily preserved. When considered together with the Pop III-like abundance pattern of the absorber, this environment supports the interpretation that the absorber may point to Pop III star formation in the outskirts of a mildly enriched dark matter halo.

The persistence of Pop III stars at relatively low redshifts is predicted by both cosmological simulations\cite{Johnson_08,Venditti_23a,Zier_25} and semi-analytical models\cite{Visbal_20,Liu_25b,Ventura_25}, which indicate that Pop III star formation can persist well into the epoch of reionization (EoR), down to redshifts $z \sim 6-10$. Theoretical studies describe multiple formation pathways for Pop III stars. The standard channel occurs in low-mass minihalos with low virial temperatures. The very first stars formed in these low-mass halos are free of radiation backgrounds from earlier stellar generations. Their formation relies primarily on molecular hydrogen (H$_2$) cooling. Due to the lack of a dipole moment, H$_2$ is an inefficient coolant at temperatures below $\sim200$ K, resulting in relatively spherical gas collapse and limited fragmentation. 
This process could potentially produce more massive stars than those forming today, while models also allow the formation of lower-mass stars\cite{Klessen_23}.

A plausible scenario that supports our observation is Pop III formation influenced by ultraviolet radiation backgrounds\cite{Clark_11}. Theoretical studies suggest that at lower redshifts, Pop III stars may continue to form in more massive dark matter halos\cite{Pallottini_14,Visbal_20}, including systems that have already experienced previous episodes of star formation\cite{Liu_20,Venditti_25}. 
% These halos have virial temperatures exceeding $T_{\rm vir} \sim 8000~\mathrm{K}$\cite{Becerra_15}. 
In such environments, H$_2$ cooling can initially be suppressed by strong Lyman-Werner (LW) radiation from earlier generations of stars, delaying star formation until the gas becomes sufficiently dense to be self-shielded. As a result, halos continue to accrete gas until reaching the atomic-cooling threshold, enabling efficient cooling via Lyman-$\alpha$ emission. 
Pop III formation at large halo masses is one of the possible scenarios to explain the bright Pop III candidates during the EoR\cite{Venditti_25}.

This extended Pop III formation window is facilitated by the patchy and inhomogeneous nature of metal enrichment, which may allow chemically pristine gas pockets to survive even within globally enriched halos, particularly in models where the metal mixing is inefficient\cite{Sarmento_22}. Furthermore, halos experiencing delayed collapse due to LW suppression or photoionization feedback may reach atomic-cooling conditions later in cosmic time, contributing to Pop III formation. Collectively, these mechanisms support a prolonged and spatially diverse Pop III formation history beyond the initial onset at Cosmic Dawn and down to $z\sim6$.

Atomic-cooling halos with masses of $\log(M_{\mathrm{h}}/M_{\odot})\gtrsim9$ are predicted to be a major site of Pop III star formation, since LW radiation is less effective in photodissociating $\text{H}_2$ molecules in more massive dark matter halos due to efficient cooling\cite{Liu_20}. 
Given our measured lower limit on the halo mass associated with the absorber, $\log(M_{\mathrm{h,min}}/M_{\odot}) > 9$, the absorber is likely to trace gas in an atomic-cooling halo. This is consistent with a scenario in which the Pop III-enriched gas is preserved in the outskirts of relatively massive, mildly enriched halos hosting more evolved Pop \textsc{ii} and/or Pop \textsc{i} stellar populations (Fig. 3).
In the Ly$\alpha$ forest from the quasar spectrum, we observe two transmission peaks near the absorber redshift (Supplementary Figure \ref{fig:transmission}).
Since the immediate absorber and host galaxy environment is likely associated with a gas overdensity that enhances the Ly$\alpha$ optical depth and suppresses transmission locally\cite{Kakiichi_25}, the transmission maxima are not expected to coincide with the exact location of the galaxy overdensity. Instead, the Ly$\alpha$ and Ly$\beta$ peaks suggest that the absorber is embedded in a larger-scale ionized region, indicating that its surrounding environment is exposed to radiation. This provides environmental context compatible with the theoretical Pop III formation pathway in the presence of the UV background \cite{Jeong_26}.

\begin{figure}[!t]
    \centering
      \includegraphics[width=1\linewidth]{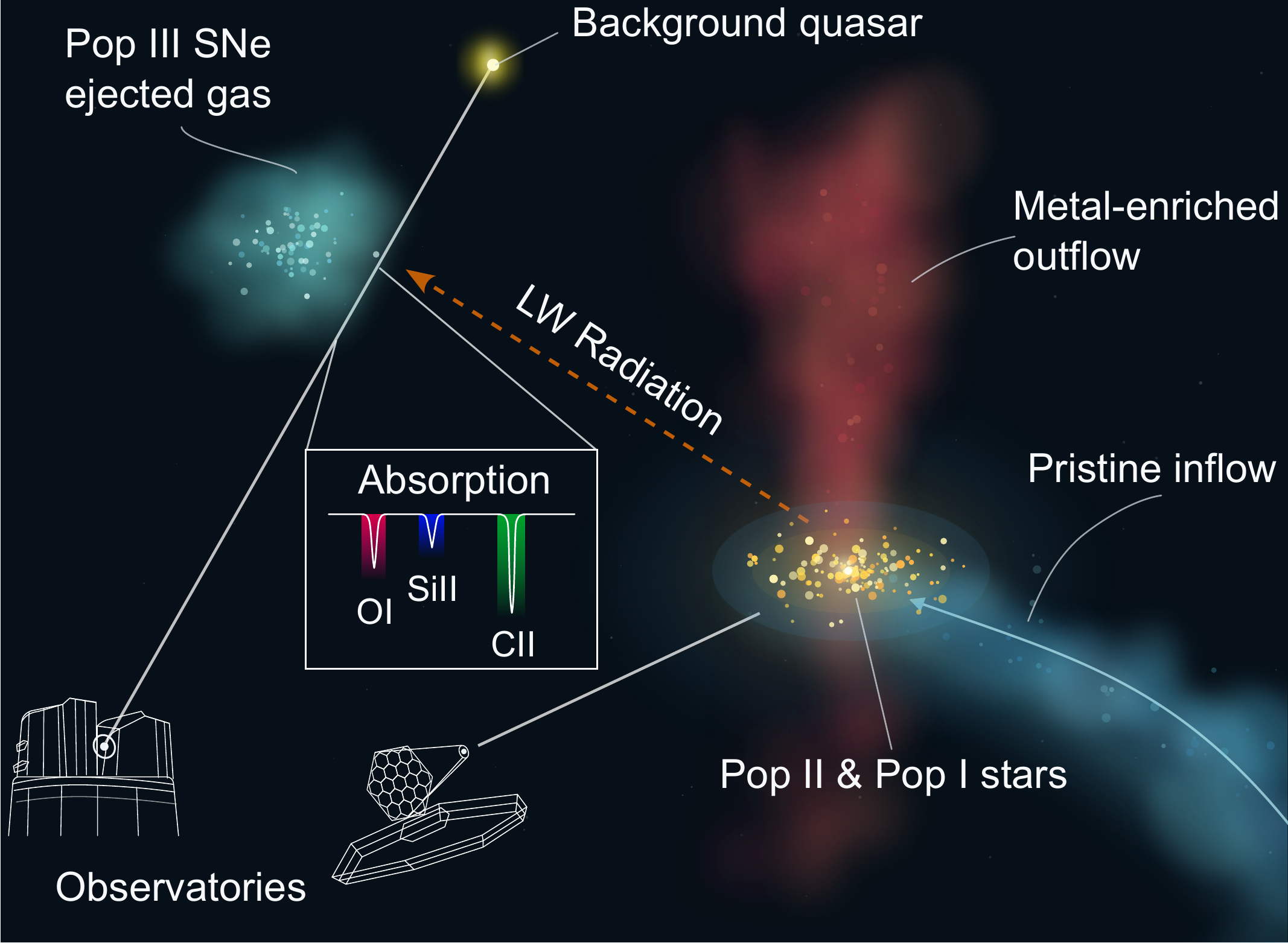}
    \caption{\small Schematic representation of the delayed Pop III formation scenario consistent with our observations. This channel occurs in atomic-cooling dark matter halos fed by primordial gas, where Pop III stars coexist with more evolved stellar populations that produce the ambient UV background, which delays Pop III star formation. The faint Pop III-forming gas clumps are likely spatially separated from the central galaxy, allowing them to remain largely free of metal pollution. Supernovae yields from the Pop III stars enrich the surrounding gas, producing distinctive absorption features in the spectrum of the background quasar. JWST can detect the brighter central galaxy, whereas the faint Pop III-forming clump remains below the emission-line detection limit and can instead be detected through absorption spectroscopy, highlighting the complementary nature of the two observational approaches.
    \label{fig:cartoon}}
\end{figure}

Alternatively, radiation-hydrodynamical simulations from \textsc{thesan-zoom} suggest that Pop III star formation can continue in low-mass haloes until the end of reionization, around $z\sim5$\cite{Zier_25}. In this picture, Pop III stars can form in low-mass subhaloes in the environments of more massive systems, often outside the virial radius of the central halo, before these subhaloes are later accreted through hierarchical assembly. Although the short-lived massive Pop III stars themselves would have quickly ended their lives, their SN ejecta may remain chemically distinguishable if they avoid substantial mixing with metals produced by subsequent Pop \textsc{ii} supernovae. As hierarchical merging proceeds, the minihaloes can be incorporated into central systems, allowing Pop III relics to persist in the outskirts of relatively massive dark matter haloes at lower redshifts. By the end of reionization, photoevaporation may suppress further star formation in these minihaloes.

In our observations, the nearest \OIII emitter to the absorber has a three-dimensional distance of 238 pkpc with an impact parameter of 119 pkpc. This distance represents $\sim14^{+34}_{-7}~r_{\rm vir}$, based on our estimate of the virial radius inferred for the absorber-associated dark-matter halo of $r_{\rm vir}=17^{+18}_{-12}$ pkpc (Methods).
In the literature, the majority of metal-enriched neutral gas at $z > 5$, produced by nearby stellar populations, is predicted to reside beyond the virial radius\cite{Pruto_25}, where it can be observed as neutral gas absorbers such as O\textsc{i}. 
Recent JWST observations have shown that metal absorbers are found remarkably far from galaxies, indicating that metal absorbers such as Mg\,\textsc{ii} and O\,\textsc{i} are frequently detected at high impact parameters as high as several times the virial radii\cite{Bordoloi_24,Higginson_25}. They have also shown that absorbers associated with galaxies at $z\sim 6$ exhibit higher [C/O] abundances compared with those lacking galaxy associations, which may indicate enhanced contributions from Pop III nucleosynthesis\cite{Higginson_25}.
Therefore, our observation is consistent with two primary scenarios for Pop III formation in the outskirts of massive dark matter halos: (1) formation in pristine regions of atomic-cooling halos, where Pop III star formation is delayed by UV background, or (2) early Pop III formation within pristine low-mass satellite halos that later merge into the central massive halo.

Our findings demonstrate that Pop III-enriched absorption features persist in the outskirts of dark matter halos at $z \sim 6$, specifically within metal-poor overdense environments. The discovery of such a system at $z = 5.945$ suggests that Pop III star formation may have continued well into the EoR, and that pristine gas reservoirs in the form of minihalos or gas clumps can survive in the external galactic regions as predicted by simulations\cite{Tornatore_07,Venditti_23a,Zier_25}.
The metal deficiency of this galaxy overdensity implies that the key difference likely lies in its chemical enrichment history, such as delayed enrichment, inefficient metal mixing, or sustained accretion of pristine gas. This metal-deficient environment therefore points to either a large, coherent region that has remained only weakly enriched or a primordial environment that is continuously replenished by strong inflows of pristine gas. In either case, the low metallicity is plausibly linked to a larger reservoir of pristine gas, which favors late-time Pop III star formation. Moreover, recent simulations\cite{Jeon_26} suggest that the overdense environments also offer a complementary pathway for identifying Pop III relics alongside more typical normal-to-underdense environments.

In summary, we use the metal absorbers as a diagnostic signpost to identify potential Pop III environments. We reveal a very low-metallicity galaxy overdensity at $z \sim 6$, where continuous replenishment by pristine gas or inefficient internal metal mixing may have allowed both pristine gas and Pop III-enriched gas to persist. These conditions provide a natural physical context for theoretical scenarios in which Pop III star formation might persist until $z \sim 6$ in the external galactic regions around relatively massive galaxies.
This encompasses a range of high-impact-parameter environments, including satellite subhalos\cite{Crosby_13} and infalling pristine gas clumps\cite{Tornatore_07,Venditti_24}. These peripheral regions are less affected by global metal enrichment, allowing Pop III stars to coexist with more evolved stellar populations before eventually being integrated into the massive host via hierarchical merging. 
In this context, our findings are consistent with scenarios in which sporadic episodes of Pop III star formation, or the survival of Pop III-enriched gas, may occur even after Pop \textsc{ii} star formation has become dominant, as found in simulations\cite{Venditti_23a,Zier_25}. Although the absorbers are not expected to show direct Pop III stellar or nebular emission, they may preserve the chemical imprint of earlier Pop III supernova enrichment. Searching for such enrichment signatures in metal-absorbing systems therefore provides a complementary avenue for probing the legacy of Pop III stars.
The prolonged survival and eventual death of Pop III stars mark an extended era spanning the first billion years of cosmic history. Accordingly, the processes observed in this overdensity likely reflect fundamental aspects of hierarchical structure formation, preserved in the fossil record of present-day galaxies\cite{Lian_23}. Finally, linking Pop III-enriched absorbers with line-emitting counterparts provides a promising framework for identifying analogous environments and establishing a physical connection between absorber-based chemical diagnostics and the galaxies embedded within or surrounding these structures.

%%%%%%%%%%%%%%%%%%%%%%%%%%%%%
%%%%%%%%%%%%%%%%%%%%%%%%%%%%%
%			METHODS			%
%%%%%%%%%%%%%%%%%%%%%%%%%%%%%
%%%%%%%%%%%%%%%%%%%%%%%%%%%%%

% \clearpage
% \newpage

\begin{methods}

% \setcounter{figure}{0}

% \makeatletter
% \renewcommand{\thefigure}{\arabic{figure}}
% \renewcommand{\fnum@figure}{Supplementary Figure \thefigure}
% \renewcommand{\theHfigure}{ED\arabic{figure}} % if hyperref is used
% \makeatother

\subsection{Cosmology and conventions}
Throughout this article, we adopt the AB magnitude system, and Planck 2018 cosmology\cite{Planck_18} with $\Omega_m=0.3097,\Omega_\Lambda=0.6888,\Omega_b=0.0490,h=0.6732$.
Emission lines are indicated as follows: $\OIII\lambda5008$ := \OIII, $\OII \lambda\lambda3727, 3730$ := \OII, if presented without wavelength values. Physical and comoving distances are distinguished by the prefixes p and c, respectively (e.g., pMpc/pkpc and cMpc/ckpc).

\subsection{Quasar spectroscopy}
The quasars SDSS J0100+2802 and VDES J0224-4711 are part of the ``XQR-30" survey\cite{dodorico23}, an ESO Large Program (ID: 1103.A-0817; P.I. V. D’Odorico) that obtained 248 hours of observations for quasars at $z\geq5.8$ with brightest J magnitude using the X-shooter spectrograph\cite{Vernet_11} on the Very Large Telescope (VLT). Optical spectra (VIS, $\lambda = 5500-10250~\AA$) were acquired with a $0\farcs9$ slit, corresponding to a nominal resolving power of $R_{\mathrm{VIS}} \sim 8900$. Near-infrared spectra (NIR, $\lambda = 9800-24800~\AA$) were obtained using a $0\farcs6$ slit, yielding a resolving power of $R_{\mathrm{NIR}} \sim 8100$\cite{sodini24}. The data reduction is thoroughly described in ref.~\citen{dodorico23}, and the general absorber catalog is provided by ref.~\citen{Davies_23}. The SDSS J0100+2802 quasar is additionally observed with Magellan/FIRE and Keck HIRES/MOSFIRE\cite{Higginson_25}, and the metal abundance ratios from the added high-resolution spectroscopy are consistent with the XQR-30 catalog.
\subsection{JWST observations and data reduction}
We utilize the NIRCam F356W grism observations targeting the luminous quasar SDSS J0100+2802 from the EIGER survey and quasar VDES J0224-4711 from the ASPIRE survey, supplemented by direct imaging in the F356W, F200W, and F115W filters.

We calibrate the individual NIRCam WFSS and direct image exposures using version 1.13.4 of the JWST Calibration Pipeline \texttt{CALWEBB} Stage 1, with reference files \texttt{jwst\_1321.pmap}. 
The $1/f$ noise is removed along detector rows for Grism-R exposures, and along both directions for images. The world coordinate system (WCS) information is assigned to each exposure via the \texttt{assign\_wcs} step, and flat-field corrections are applied using \texttt{CALWEBB} Stage 2.

We construct median background images from all WFSS and direct image exposures, which are then scaled and subtracted from each frame. The direct images are aligned to the ninth data release of the DESI Legacy Surveys catalog (LS\_DR9) and drizzled to a common pixel scale of 0\farcs03/pixel using the Grism Redshift \& Line Analysis tool (\textsc{Grizli}\cite{Brammer_2022}).
Astrometric offsets between the SW images and the calibrated F356W mosaic are measured to align each grism exposure with the direct image, ensuring proper WCS alignment and enabling accurate spectral tracing.

The pre-processed WFSS exposures are subsequently analyzed using \textsc{Grizli}, employing the V9 spectral tracing and grism dispersion models (\url{https://github.com/npirzkal/GRISM_NIRCAM}). Detection catalogs used for spectral extraction are generated from the F356W direct image. Continuum cross-contamination is mitigated via \textsc{Grizli}'s forward modeling, using the F356W image as the reference for each grism exposure. One-dimensional spectra are extracted using the optimal extraction method\cite{Horne_86}, with extraction profiles derived from the F356W image and its corresponding segmentation map.

\subsection{Emitter catalog and photometry}
The construction of the \OIII emitter catalog in the ASPIRE survey is outlined in ref.~\citen{Wangf_23,Li_25}, and the \OIII catalog in the EIGER survey is detailed in ref.~\citen{Kashino_23,Kashino_25}. In brief, the emission line candidates are identified by applying a peak-finding algorithm on median-filter spectra, followed by Gaussian fitting and selection based on line signal-to-noise ratio. Assuming the detected line is $\rm \OIII\lambda5007$, we verified the corresponding $\rm \OIII\lambda4959$ line, applied a significance cut, and finally conducted visual inspection to reject likely misidentified sources.

The photometric catalog is constructed in the same way as ref.~\citen{Li_25}. The sources are first identified using F356W as the detection image. Photometry is measured using \texttt{SourceXtractor++} in F115W, F200W, and F356W after matching the PSFs to those of F356W (0\farcs1). Fluxes measured in Kron apertures with ($k$=1.2, $R$=1.7) are corrected to total magnitudes using the ratio for Kron fluxes measured in larger apertures ($k=1.5, R=2.5$). The uncertainties for each source's photometry are measured by placing 1000 random apertures the size of the Kron aperture across the image and measuring the root mean square (RMS).

\subsection{Absorber-galaxy sample selection}
We started from the XQR-30 metal-absorber catalog and required detections of \CII, O~\textsc{i}, and \SiII in order to derive robust [C/O] and [Si/O] abundance ratios. We compared these ratios with Pop III enrichment models (Fig. 2a) and cross-matched the corresponding absorbers with quasar sightlines covered by JWST/NIRCam grism observations from ASPIRE and EIGER. The absorber at $z=5.945$ toward SDSS J0100+2802 is the only system in this set with abundance ratios consistent with Pop III enrichment, and its JWST data reveal an associated galaxy overdensity. 
To place this system in context, we aimed to select a comparison absorber from the same parent sample with similar observational information but without Pop III\-enriched abundances. After excluding the $z=6.06$ absorber toward PSO J036+03, which is associated with a single isolated galaxy\cite{Zou_24}, and the $z=6.11$ and $z=6.14$ absorbers toward SDSS J0100+2802, for which no associated galaxies are identified\cite{Kashino_25}, the final sample consists of two absorbers associated with galaxy overdensities: the Pop III-enriched system toward SDSS J0100+2802 and the non-Pop III comparison system toward VDES J0224-4711.
\subsection{Overdensity characterization}
\begin{suppfigure}[!b]
    \centering
      \includegraphics[width=1\linewidth]{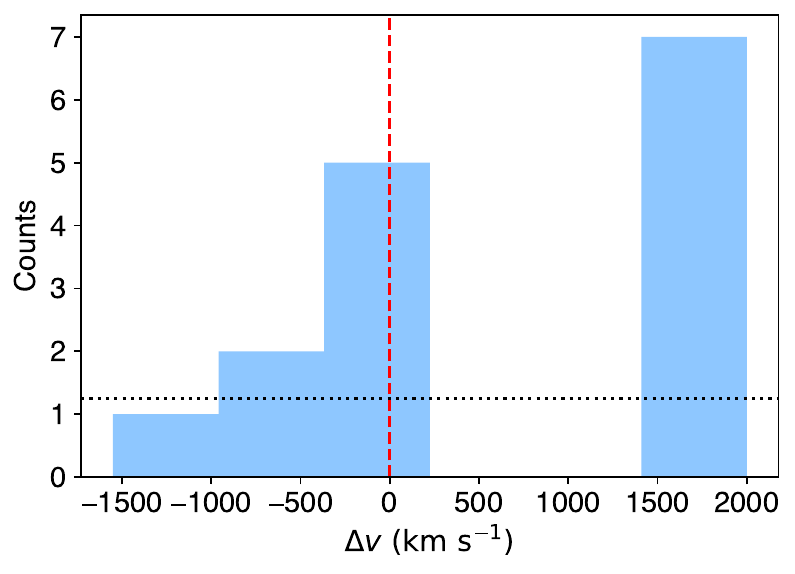}
    \caption{\small The velocity distribution of the 17 galaxies in overdensity in the J0100+2802 field. The absorber redshift $z=5.945$ is taken as the velocity zero point, marked by a red dashed line. The black dotted line marks the mean number density of the whole field.
    \label{fig:dv}}
\end{suppfigure}
Using the coordinates and redshifts of each galaxy, we identify the overdense structures with the FoF clustering algorithm\cite {Li_25}. This method has been widely applied to dark matter halos\cite{More_11}, and here we apply it to galaxies.
This algorithm selects galaxy groups by searching for companions around a central galaxy within a projected separation $d_{\text{link}}=500\ \rm{pkpc}$ and line-of-sight (LOS) velocity $\sigma_{\text{link}}=500\ \rm{km/s}$. The algorithm iteratively performs the process above for all the companions identified until no more galaxies are found. This process selected two separate groups near the absorber, with mean redshifts $z\approx5.94$ and $z\approx5.99$. The velocity distribution of the two groups is shown in Supplementary Figure \ref{fig:transmission}. Using the gapper scale estimator\cite{Beers_90}, which is recommended for samples with small numbers, we obtain a velocity dispersion of $1253 \pm 133~\rm km~s^{-1}$ for all these galaxies.

Previous observations of galaxy overdensities at redshifts $z\gtrsim6$ show large line-of-sight velocity dispersions ($\sigma\sim1000~\rm km~s^{-1}$)\cite{Morishita_23}, and simulations demonstrate that proto-clusters at these epochs are spatially extended across tens of cMpc\cite{Chiang_13}. Within such dynamically young and less virialized systems, two overdense groups separated by $\sim 1000~\rm km~s^{-1}$ can therefore be plausibly interpreted as substructures embedded in the same forming large-scale environment. Indeed, adopting a more permissive linking-velocity threshold of $\sigma_{\text{link}} = 1500~\rm km,s^{-1}$ results in all identified galaxies being associated with a single structure comprising 17 members. We therefore present the characterization of the absorber and its connection to the surrounding large-scale environment of these 17 galaxies.

\subsection{Galaxy-IGM connection}
\begin{suppfigure}[!b]
    \centering
      \includegraphics[width=1\linewidth]{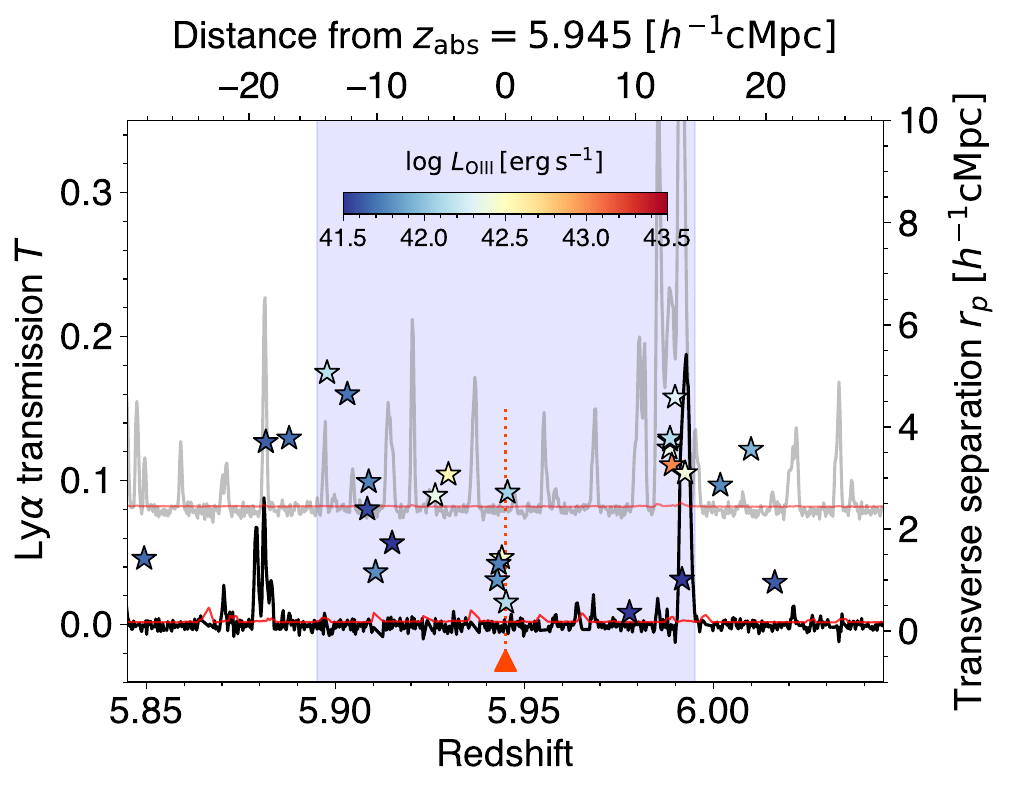}
    \caption{\small Ly$\alpha$ (black line) and Ly$\beta$ (gray line) forest transmission $T=\exp(-\tau_\alpha)$ with uncertainty (red line) around the Pop III absorber $z_\text{abs}=5.945$. The Ly$\beta$ transmission is shifted upwards. The red triangle and vertical dotted line mark the redshift of the Pop III absorber. The locations of \OIII emitters relative to the absorber are marked by stars, color-coded by \OIII luminosity.  The top $x$-axis indicates the line-of-sight comoving distance from the absorber, and the right $y$-axis indicates the transverse distance of \OIII emitters from the absorber. The blue shaded region indicates the redshift interval used to calculate the \OIII clustering around the absorber.} \label{fig:transmission}
\end{suppfigure}
We show the Ly$\alpha$ and Ly$\beta$ forest transmission and the distribution of \OIII emitters centered around the absorber in Supplementary Figure \ref{fig:transmission}. We observe two notable transmission spikes at $z \approx 5.99$ and $z \approx 5.88$, located at $\sim-20$, and $\sim16$ $h^{-1}\rm cMpc$ from the absorber, which were also reported by ref.~\citen{Kashino_25}. The offset from the absorber redshift is expected\cite{Meyer_19} as the Ly$\alpha$ forest transmission at the immediate vicinity of absorber-host galaxies is dominated by gas overdensities\cite{Garaldi_24}.  
Ref.~\citen{Kakiichi_25} indicates an excess of Ly$\alpha$ transmission at $z\sim6$ around \OIII emitters on a scale of $\sim20-40$ cMpc, consistent with the separation of the two Ly$\alpha$ transmission spikes relative to the absorber redshift. The presence of Ly$\beta$ spikes besides the Ly$\alpha$ spikes is also supportive of the single ionized region.
Thus, these spikes are indicative of a localized ionized bubble in the region of $\gtrsim36\,h^{-1}\rm\,cMpc$ in size, spatially coincident with an overdensity of \OIII-emitting galaxies. 

We estimate the H\textsc{i} photoionization rate by Lyman-$\alpha$ transmission $T$ at the redshift of the absorber in the quasar spectrum. 
We calculate the observed mean Lyman-$\alpha$ transmission in the redshift range within the redshift interval of $\pm0.05$ from the Pop III absorber, identical to the range we chose to calculate the cross-correlation function. We obtain the mean transmission around Pop III absorber as $T_\text{obs}=(2.67\pm0.10)\times10^{-3}$, corresponding to observed optical depth $\tau_\text{obs}=5.93\pm0.04$, which is comparable to the Ly$\alpha$ optical depth measured around \OIII emitters at similar redshifts\cite{Jin_24}.

The mean Lyman-$\alpha$ transmission can be theoretically expressed as the average of the local transmission over the baryonic density field:
\begin{equation}
    T=\langle e^{-\tau_\alpha} \rangle = \int d\Delta_b P_V(\Delta_b) e^{-\tau_\alpha},
\end{equation}
where $P_V$ is the volume-weighted density PDF, $\Delta_b$ is the baryonic overdensity, and $\tau_\alpha$ is the optical depth.
Using the fluctuating Gunn-Peterson approximation, the optical depth is related to H\textsc{i} photoionization rate\cite{Becker_15}:
\begin{equation}
    \tau_\alpha \simeq 10.8 \Delta_b^2 \left( \frac{\Gamma_{\text{H}\textsc{i}}}{10^{-12} \, \text{s}^{-1}} \right)^{-1} \left( \frac{T_0}{10^{4} \, \text{K}} \right)^{-0.72} \left( \frac{1 + z}{7} \right)^{9/2}.
\end{equation}\label{eq:GP}
Assuming the local IGM gas temperature $T_0=10^4$ K, solving the above equations with the observed transmission, we obtain $\Gamma_{\text{H}\textsc{i}}=(0.167\pm0.002)\times10^{-12}~s^{-1}$. Given that the gas temperature is unconstrained, we calculate $\Gamma_{\text{H}\textsc{i}}=(0.102\pm0.002)\times10^{-12}~s^{-1}$ if the gas is ionized to a higher temperature $T_0=2\times10^4$ K\cite{Davies_24}. We therefore adopt a more conservative range for the photoionization rate, $\Gamma_{\text{H}\textsc{i}} = (0.100-0.169)\times10^{-12}~\rm s^{-1}$, accounting for uncertainties in both the transmission spectrum and the gas temperature.
The measured $\Gamma_{\text{H}\textsc{i}}$ around the absorber is also consistent with the mean photoionization rate at $z\sim6$ obtained from recent analyses of large quasar spectra samples\cite{Gaikwad_23,Davies_24}. However, the Ly$\alpha$ transmission is mainly boosted by the transmission spikes; we further estimate the local ionization background on the scale of the mean free path. We apply the ionizing photon mean free path $\lambda_\text{mfp}=24.5$ cMpc at $z=5.9$\cite{Davies_24} (see also ref. \citen{Becker_21,Zhu_23}). Since we did not detect transmission signal within the scale $\lambda_\text{mfp}$, we estimate the $5\sigma$ upper limit on the photoionization rate $\Gamma_{\text{H}\textsc{i}}<0.11\times10^{-12}~s^{-1}$, at temperature $T_0=10^4$ K, indicating that the absorber is irradiated by moderate or low UV background. 
% Compare with ref~\citen{Haardt_12}

The coexistence of the ionized bubble indicates a permeating UV and LW background in the overdense environment, making the region susceptible to late-time Pop III formation. In such environments, the external radiation field suppresses H$_2$ cooling in minihalos while Pop III stars subsequently form in more massive atomic-cooling halos where efficient atomic hydrogen cooling enables gas cloud collapse despite external radiation. 
These results support the presence of low-metallicity star formation in a chemically primitive environment, which may also provide favorable conditions for theoretical scenarios in which Pop III star formation is delayed by LW radiation\cite{Jeong_26}.

\subsection{Chemical enrichment model}
To uncover the chemical signature of Pop III SNe, we use the simple and general parametric study first introduced by ref.~\citen{Salvadori_19} and then generalized by ref.~\citen{Vanni_23, Vanni_24}. The model predicts the chemical abundances (elements from C to Zn) of gaseous environments imprinted by Pop III SNe only and by both Pop III SNe and subsequent generations of normal Pop II stars. Specifically, we adopt a critical metallicity of $Z_{\rm cr}=10^{-4.5}Z_\odot$ to distinguish Pop III from Pop II star formation\cite{Vanni_23}. The adopted solar metallicity is $Z_\odot=0.0126$ in absolute mass fraction\cite{Asplund_09}.
Thus, given a set of observed chemical abundance ratios (both in metal-poor gas or stars), the model is able to explore the contribution of Pop III SNe to its metal-enrichment and to uncover the properties of the Pop III SNe, such as their progenitor mass and explosion energies, under the adopted assumptions\cite{Vanni_24}.

To achieve this goal, the model performs a parametric exploration, quantifying the unknowns related to early cosmic star formation in three free parameters: the star formation efficiency, $f_*$, which quantifies the amount of cold gas turned into stars; the dilution factor, $f_{dil}$, which establishes the amount of gas available to dilute metals; the fraction of Pop III enrichment, $f_{\rm Pop~\textsc{iii}}$, which quantifies the amount of metals in the gas provided by Pop III SNe with respect to the total metal content.

The chemical abundances of the gas, [X/H], depend on the ratio $f_*/f_{dil}$ and the fraction of Pop III enrichment, $f_{\rm Pop~\textsc{iii}}$\cite{Vanni_23}.
Thus, these quantities are evaluated by exploring the entire parameter space: $f_*/f_{dil}\in[10^{-4}-10^{-1}]$ and $f_{\rm Pop~\textsc{iii}}\in[1,0.01]$\cite{Salvadori_19}. In contrast, the relative abundance ratios between different chemical elements, [X/Y], are primarily sensitive to $f_{\rm Pop~\textsc{iii}}$, while their dependence on $f_*/f_{dil}$ is only indirect, setting the initial metallicity of subsequent generations of Pop II stars. The model employs the Pop III SN yields\cite{heger02,heger10}, which represent the most complete set of yields available in the literature. This provides a broad grid of Pop III supernova yields spanning progenitor masses of ($10-100$ \Msun and $140-260$ \Msun) and SN explosion energy of Pop III stars (from low-energy SNe to hypernovae and PISNe). For normal Pop II stars, the model assumes the yields of ref.~\citen{Limongi_18} (non-rotating, set R).

The model has been extensively tested against observations of ancient metal-poor stars\cite{Vanni_23,Skuladottir_24b} and distant metal absorbers\cite{Vanni_24,sodini24}. In particular, the abundances of some extremely metal-poor stars are consistent with enrichment by individual low-energy, or faint, and high-energy Pop III supernovae \cite{Iwamoto_05,Placco_11,Keller_14,Frebel_15,Skuladottir_24}. Independent cosmological chemical-evolution models and simulations of the Milky Way halo and ultra-faint dwarf galaxies also recover similar ranges of Pop III progenitor masses and explosion energies\cite{Salvadori_14,Cooke_14,Ishigaki_18,Koutsouridou_23,Rossi_25}, providing an independent consistency check on the parameter ranges explored here.

At the high-mass end, recent observations of a candidate Pop III stellar clump at $z=10.6$ (Hebe\cite{Maiolino_26}) suggest the possible presence of very massive zero-metallicity stars. This is also supported by theoretical work indicating that the Pop III mass spectrum may extend beyond 250 $M_\odot$ \cite{Rusta_26}. These indirect lines of evidence motivate the inclusion of the pair-instability mass range in our model grid. We therefore interpret the inferred Pop III-like signatures not as a unique solution, but as features that remain consistent across a broad and observationally motivated range of Pop III enrichment scenarios.

Though these consistent results demonstrate the robustness of the model, we caution that Pop III nucleosynthesis models remain uncertain, as the first stars and their explosions have not been directly observed. 
Other enrichment channels may also contribute to parts of the observed pattern, but they are less likely to explain the combined C/O and Si/O enhancement on their own. AGB stars mainly produce C, N, Na, Mg, and $s$-process material, and evolve on longer timescales. They are more relevant to explain the enhanced N/O patterns, such as those discussed for GN-z11 \cite{Antona_23,Cameron_23}. Similarly, very metal-poor fast rotators\cite{Meynet_06}, supermassive stars\cite{Charbonnel_23}, and Wolf--Rayet channels can provide CNO-rich material and may contribute to the carbon enhancement, but they do not naturally release large amounts of Si through winds alone. Since Si is mainly synthesized in deeper layers and/or during the supernova phase, the simultaneous enhancement of C/O and Si/O in this Pop III-like absorber is more naturally explained by a supernova contribution. We therefore interpret individual abundance patterns cautiously, focusing not on a unique diagnostic solution but on whether the inferred Pop III-like signatures remain consistent with Pop III enrichment across the explored parameter space.

\subsection{Absorber metal abundance}

We measure the metal abundance ratios using the publicly released XQR-30 catalog\cite{dodorico23}. The column density measurements for each element are obtained through Voigt profile fitting of absorption systems\cite{Davies_23}. The silicon and carbon abundance are defined as $[\text{Si}/\text{O}] = \log\left( \frac{N_{\text{Si}}/N_{\text{O}}}{N_{\text{Si}_\odot}/N_{\text{O}_\odot}} \right)$, $[\text{C}/\text{O}] = \log\left( \frac{N_{\text{C}}/N_{\text{O}}}{N_{\text{C}_\odot}/N_{\text{O}_\odot}} \right)$, where $N_{\text{Si}}/N_{\text{O}}$ and $N_{\text{C}}/N_{\text{O}}$ represent the column density ratios of silicon-to-oxygen and carbon-to-oxygen, respectively, and $\odot$ refers to the corresponding solar values. Specifically, we apply the solar reference values\cite{Asplund_21}: $\log(N_{\text{Si}_\odot}/N_{\text{O}_\odot})=-1.18$ and $\log(N_{\text{C}_\odot}/N_{\text{O}_\odot})=-0.23$. 

While the high ionization lines (C\textsc{iv} and Si\textsc{iv}) in the Pop III absorber candidate were previously reported\cite{sodini24,Davies_23}, they have a velocity offset of $66\rm~km~s^{-1}$ with respect to the low-ionization lines, and the Doppler parameter shows that they have a different velocity dispersion. 
For J0100+2802, the VIS-arm resolving power is $R \simeq 11400$, corresponding to ${\rm FWHM} \simeq 26~{\rm km~s^{-1}}$ and $b_{\rm instr} \simeq 16~{\rm km~s^{-1}}$\cite{dodorico23}. The absorption features are therefore sufficiently resolved, allowing lines with intrinsic $b \sim 5-10~{\rm km~s^{-1}}$ to be distinguished.
The velocity offset tentatively indicates that the high-ionization lines may not be physically associated with the gas cloud as the low-ionization lines. 
To be conservative, we estimate the $3\sigma$ upper limit for C\textsc{iv} and Si\textsc{iv} at the region with the same velocity as low ionization lines, and get $\log(N_{\text{C}\textsc{iv}}/\rm{cm^{-2}})<11.8$ and  $\log (N_{\text{Si}\textsc{iv}}/\rm{cm^{-2}})<11.4$. 

The abundances are reported in Table \ref{tab:ab}. We note that the measured abundances retrieved from the public XQR-30 catalog\cite{Davies_23} show slight differences from those reported by ref. \cite{sodini24}, who carried out independent detection and fitting procedures, and their results are consistent with the official catalog within $1\sigma$.

\subsection{Ionization correction}

\begin{suppfigure*}[t!]
\centering
  \includegraphics[width=1\linewidth]{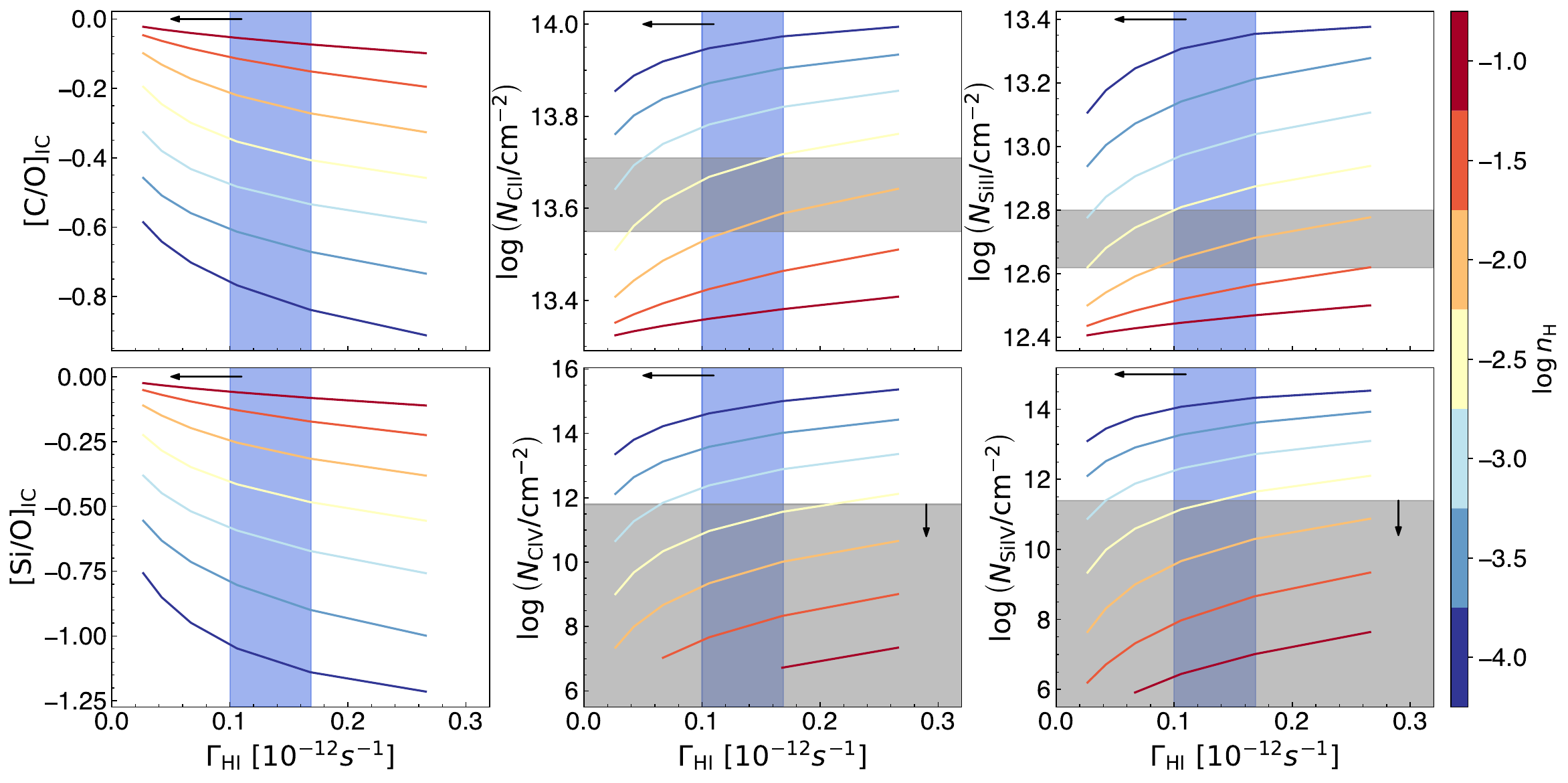}
\caption{\small The \textsc{Cloudy} model outputs at $\log(N_{\text{ H}\textsc{i}}/\rm cm^{-2})=19.15$. The first column shows the ionization correction for [C/O] and [Si/O]; the second and third columns show the predicted column densities for C\textsc{ii}, C\textsc{iv}, Si\textsc{ii}, and Si\textsc{iv}. The lines represent different models, color-coded by hydrogen density $\log n_\text{H}$. The $1\sigma$ interval of the observed column densities for C\textsc{ii} and Si\textsc{ii} are marked by gray shadows, and the $3\sigma$ upper limits for C\textsc{iv} and Si\textsc{iv} are marked by gray shadows with a black arrow. The $\Gamma_{\text{H}\textsc{i}}$ estimated within the redshift interval $z \pm 0.05$ is indicated by blue vertical shaded regions, while the $5\sigma$ upper limits measured within the mean free path are denoted by black arrows pointing to the left. We note that a lower correction is needed if we adopt a higher $\log(N_{\text{H}\textsc{i}})>19.15$.}\label{fig:IC}
\end{suppfigure*}

The ionization correction for metal absorbers at $z\sim6$ is believed to be small, because O\textsc{i} is highly coupled with H\textsc{i} and traces dense, neutral gas\cite{Becker_19,Doughty_19}, which provides a condition for self-shielding from the external UV background. In addition, the high ionization lines, such as C\textsc{iv} and Si\textsc{iv}, are not detected at the same velocity as this O\textsc{i} absorber\cite{Davies_23}, suggesting that the gas pocket remains largely neutral. Thus, previous works did not apply ionization corrections for $z\sim6$ absorbers and assumed that the low-ionization lines (i.e., O\textsc{i} and C\textsc{ii}) trace the corresponding total metal abundance\cite{christensen23,sodini24,Higginson_25}.

However, for a stricter analysis, we further constrain the ionization corrections for the Pop III metal absorber using \textsc{Cloudy}\cite{Gunasekera_25} models.
The ionization correction is defined as the deficit between the intrinsic ($\log\mathrm{N_{int}}$) and observed ($\log\mathrm{N_{obs}}$) column density:
\begin{equation}
    \mathrm{IC}=\log\mathrm{N_{int}}-\log\mathrm{N_{obs}}
\end{equation}
Following ref.~\citen{Berg_25}, we constructed \textsc{Cloudy} model grids by sampling parameters across the redshift range $5.40 \leq z \leq 6.45$, H\textsc{i} column densities $17 \leq \log N(\mathrm{H}\textsc{i}) \leq 20.5$, volume density of hydrogen $-4\leq n_\text{H}\leq0$, metallicities $-3.9 \leq [\mathrm{M}/\mathrm{H}] \leq -1.5$, and varying photoionization rates assuming the spectral shape of the Haardt \& Madau UV background model\cite{Haardt_12}. We assumed an intrinsic solar abundance pattern, but the carbon and silicon abundances are elevated by 0.3 dex to simulate the level of Pop III enrichment expected by theoretical models\cite{Vanni_24}.
The AGN contribution to the radiation field was neglected, as none of the absorbers in this study are located proximate to quasars. The predicted column densities of each element at given physical conditions are obtained by interpolating the model results across these parameter grids.
The final \textsc{Cloudy} grid varies four parameters $\log N(\mathrm{H}\textsc{i})$, $n_\text{H}$, $\rm [X/H]$, and $\Gamma_{\text{H}\textsc{i}}$.

The ionization correction is obtained by comparing Cloudy model parameters to observational data, ensuring that the predicted line ratios are consistent with the observations\cite{Berg_25}. Since the H\textsc{i} column densities cannot be measured directly for our absorber, we estimate the possible ionization corrections by exploring a range of plausible H\textsc{i} column densities. Ref~\citen{sodini24} estimated HI column densities indirectly by exploiting the observed relation between O\textsc{i} and H\textsc{i} in very metal-poor DLAs\cite{Cooke_11}.
By assuming $\rm [O/Fe] \approx 0.3$ produced by Type II SNe, which is also applicable for Pop III SNe enrichment in the mixture model\cite{Vanni_23}, they found the empirical relation between $\rm N(O\textsc{i})$ and $\rm N(H\textsc{i})$:
\begin{equation}
\rm \log N(O\textsc{i}) = log N(H\textsc{i}) + \log(n_O/n_H)_\odot + [O/Fe] + [Fe/H]
\end{equation}
This relation depends on the metallicity of the absorption system in terms of [Fe/H]. 
The CGM metallicity can be predicted by the mass-metallicity relation of absorption systems\cite{Moller_13}. Because the gravitational mass of an absorption host halo influences the range of velocities of its absorption components, the resulting absorption line width is used as a proxy for the host's mass. The line width of low-ionization lines in our system is measured to be 11 $\rm km~s^{-1}$. Assuming that the redshift evolution of the mass–metallicity relation from $z=0$ to $z=5$ reported by ref.~\citen{Moller_13} extends to $z>5$, we estimate a metallicity of $\rm [Fe/H]\sim -2.73$.
Given that the metallicity is expected to decrease at higher redshift, we use the absorber metallicity $\rm [Fe/H]= -2.7$ estimated from the mass-metallicity relation as an upper limit, and we get the lower limit of H\textsc{i} column density $\rm log\,N(H\textsc{i})>19.15$.

Since the resulting ionization state is sensitive to the local radiation background, we vary the model input photoionization flux, $\Gamma_{\text{H}\textsc{i}}$, to investigate its effects.
Supplementary Figure \ref{fig:IC} shows the relation between $\Gamma_{\text{H}\textsc{i}}$, $\log n_\text{H}$ and the ionization correction. We find that the non-detection of C\textsc{iv} and Si\textsc{iv} can constrain the hydrogen density to $\log n_\text{H}\gtrsim-2.5$, and the observed C\textsc{ii} and Si\textsc{ii} abundances match the model's prediction with $\log n_\text{H}\sim-2$. Assuming a hydrogen number density of $\log n_\text{H}=-2$, and using the estimated H$\textsc{i}$ photoionization background that includes Ly$\alpha$ transmission spikes, we find that the ionization corrections are [C/O]$_{\text{IC}}\gtrsim-0.3$ and [Si/O]$_{\text{IC}}\gtrsim-0.3$.
Alternatively, if we only consider the local photoionization background estimated within the mean free path ($\lambda_{\text{mfp}}=24.5$ cMpc), the ionization corrections can be lower, specifically [C/O]$_{\text{IC}}\gtrsim-0.2$ and [Si/O]$_{\text{IC}}\gtrsim-0.2$. The required correction is further reduced when adopting a higher H\textsc{i} column density, as expected for metal-poor gas.
Therefore, even after applying these corrections, the absorber remains an outlier in Fig. 2a. Its abundances are still consistent with Pop III enrichment models only if at least $30\%$ of its metals originate from Pop III stars. 

\subsection{Galaxy Metal Abundance}
\begin{extendedfigure*}[t!]
\centering
  \includegraphics[width=1\linewidth]{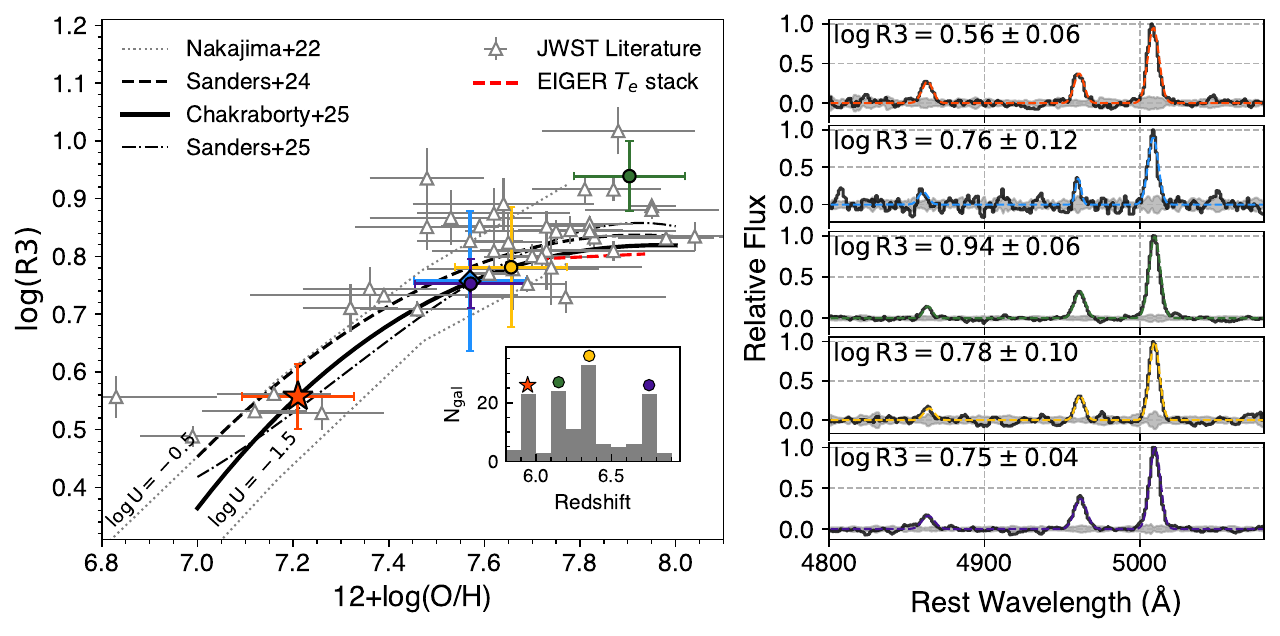}
\caption{\small {\bf Left panel}: The empirical R3 calibrations and JWST observations. Literature JWST observations with $T_e$-metallicity measurements\cite{Curti_23a,Nakajima_23,Trump_23,Jones_23,Sanders_23,Pollock_25,Cullen_25} are marked by gray triangles. The black solid line represents the fiducial empirical calibration adopted in this work (Ref.~\citen{Chakraborty_25}), whereas the black dashed line represents the empirical calibration from ref.~\citen{Sanders_23} for comparison. The black dotted curves present the photoionization model tracks with ionization parameters $\log\rm U=-0.5$ and $\log\rm U=-1.5$ from ref.~\citen{Nakajima_22}. The red dashed line marks the $T_e$-metallicity measurements obtained by stacking the \OIII emitters from the full EIGER survey\cite{Kotiwale_25}.
The red star and blue diamond indicate galaxy overdensities associated with Pop III and Pop II absorbers, respectively. The purple, yellow, and green circles represent three additional major overdensities within the SDSS J0100+2802 field, which are not associated with the Pop III absorber. The inset in the bottom-right corner displays the redshift distributions of these four overdensities, each denoted by markers with corresponding shapes and colors on top. {\bf Right panel}: Composite spectra of galaxies in each overdensity. The dashed lines represent Gaussian model fits, following the same color scheme as the markers on the left panel. The measured R3 ratios are denoted in the top-left corner of each row.} \label{fig:R3-Z}
\end{extendedfigure*}
We adopt the fiducial metallicity diagnostic presented by ref.~\citen{Chakraborty_25} to infer the gas-phase ISM metallicity of our sample galaxies. This empirical calibration is built upon a sample of $z=3-10$ galaxies with $\OIII\lambda{4363}$ detected using the direct $T_e$ method, which better characterizes the properties of high-redshift galaxies, and is also consistent with other similar calibrations\cite{Scholte_25, Sanders_23}.

To measure the average metallicity of our sample, we stack the individual spectra to construct the composite spectra. We resample our spectra to rest frame on a common 1 \AA\ wavelength grid with flux preserved using \texttt{spectres}\cite{2017arXiv170505165C}. To avoid excessive weighting toward brighter sources with strong line fluxes, we normalized each spectrum by its measured \OIII\ flux. We then computed the mean value of the normalized spectra at each wavelength bin, with uncertainties estimated from the standard deviation of 1000 bootstrap realizations of the sample. We model the \OIII\ doublets using two separate Gaussian components and fit the \Hb\ line with a single Gaussian. The \OIII/\Hb\ (R3) ratios are then measured from the best-fit Gaussian model profiles. We thus convert the R3 ratios to metallicity using empirical calibration\cite{Chakraborty_25}.

A caveat of the R3 diagnostic is that it produces a double-valued solution for metallicity\cite{Sanders_23,Sanders_25}. To resolve this degeneracy, the inclusion of the $\OII$ line allows one to distinguish between the two branches via the $\OII/\Hb$ ratio\cite{nakajima23}.
In addition, the calibration method incorporating the $\OIII/\OII$ index can account for the ionization parameter and reduce the scatter in metallicity estimates at the low-metallicity\cite{Izotov_21}. Recent works have also proposed using a linear combination of the $\OIII/\Hb$ and $\OII/\Hb$, which minimizes the scatter at fixed metallicity\cite{Laseter_24,Scholte_25,Sanders_25}. This is thought to result from reducing the secondary dependence on the ionization parameter.

However, in our sample, the \OII\ line falls outside the spectral coverage. Therefore, we adopt the lower branch solution, corresponding to $12+\log(\rm O/H) \lesssim 7.9$, based on the fact that most galaxies at comparable redshifts have been confirmed to be metal-poor through the $T_e-$metallicity method\cite{Curti_23a,Nakajima_23,Trump_23,Jones_23}. This assumption has been widely adopted in previous works\cite{Li_25,Matthee_23,Vanzella_23}. By contrast, the upper branch solution would imply super-solar metallicities given the observed R3 ratios, which are less plausible in the context of early galaxies.

In Extended Data Figure \ref{fig:R3-Z}, we show our metallicity measurements using R3 calibration\cite{Chakraborty_25}, in comparison with literature JWST $T_e-$metallicity measurements\cite{Curti_23a,Nakajima_23,Trump_23,Jones_23,Sanders_23,Pollock_25}. We find that the empirical calibrations lie well within the range of the observed data, while ref.~\citen{Sanders_23} shows $\lesssim0.1$ dex offset towards lower metallicities compared with our adopted fiducial calibration\cite{Chakraborty_25}. The photoionization models with an ionization parameter range of $\log \rm U = [-1.5, -0.5]$ from ref.~\citen{Nakajima_22} also show reasonable agreement with the literature data, suggesting hard ionizing radiation in those galaxies. The two different empirical calibrations both suggest the metal-poor nature of our sample galaxies. The systematic uncertainty between different calibrations is within $1\sigma$ statistical error of metallicity measurements. The galaxy overdensity associated with the Pop III absorber is significantly more metal-poor than other galaxy overdensities lacking Pop III signatures, regardless of the calibration used. In Extended Data Figure \ref{fig:R3-Z}, we also highlight the $T_e$-metallicity measurements from the full EIGER survey\cite{Kotiwale_25}, the parent sample of this work, with a red dashed line. While the other three overdensities are well aligned with the mean stacking results, the overdensity associated with the Pop III absorber lies significantly below the mean metallicity probed by the EIGER survey.

\subsection{The Pop III absorber-galaxy cross-correlation function}

To estimate the host halo mass of the Pop III-enriched absorber, we measure the volume-averaged projected cross-correlation function using the estimator,
\begin{equation}
    \chi_{\text{PG}}(R_{\text{min}}, R_{\text{max}}) = \frac{\langle \text{PG} \rangle}{\langle \text{PR} \rangle} - 1,
\end{equation}
where $\langle \text{PG} \rangle$ represents the number of detected galaxies within radial bins surrounding the Pop III absorber, whereas $\langle \text{PR} \rangle$ denotes the expected number of galaxies within the same volume in a randomly selected ``blank'' field. To estimate $\langle \text{PR} \rangle$, we calculate the number of absorber-galaxy pairs from our random catalog within the same radial bins and satisfying the condition $( \left|\Delta v\right| \leq v_{\text{max}} )$. We consider galaxies within the redshift interval of $\pm0.05$ from the Pop III absorber, corresponding to  the maximum velocity $\left|\Delta v\right| < v_{\text{max}}=2161\ {\rm km\ s}^{-1}$. The chosen redshift interval well encompasses the overdensity structure. These pairs are then normalized to the expected number of galaxies within the equivalent cylindrical volume in a randomly selected ``blank'' field. We estimate the expected quasar-galaxy pair counts based on the galaxy number density derived from the \OIII luminosity function\cite{Matthee_23}, multiplied by the observed volume within each shell and adjusted for the completeness of the survey volume. At large radii, the cylinder is incomplete due to the irregular footprint of the survey; we calculate the actual volume by intersecting the cylinder with the survey footprint. Uncertainties on the cross-correlation measurements are estimated by bootstrapping the galaxies and the random catalog.

The projected cross-correlation function at each radial bin between $R_{\rm min}$ and $R_{\rm max}$ can be expressed as the integral of the 3D cross-correlation function $\xi_{\rm PG}$ between Pop III absorber and galaxies along the line-of-sight,
\begin{equation}
    \chi_{\text{PG}}(r_{\text{min}}, r_{\text{max}}) = 
    \frac{2}{V} \int_{r_{\text{min}}}^{r_{\text{max}}} \int_{0}^{\pi_{\text{max}}}
    \xi_{\text{PG}}(r_p, \pi) 2\pi r_p \, \text{d}r_p \, d\pi.
\end{equation}
where $V$ is the volume of the cylindrical shell and the maximum line-of-sight comoving distance $\pi_{\rm max}=v_{\text{max}}(1+z)/H(z)=14.6\ {h^{-1}\text{cMpc}}$.

\subsection{Halo occupation distribution (HOD) model}
\begin{extendedfigure}[!t]
    \centering
      \includegraphics[width=1\linewidth]{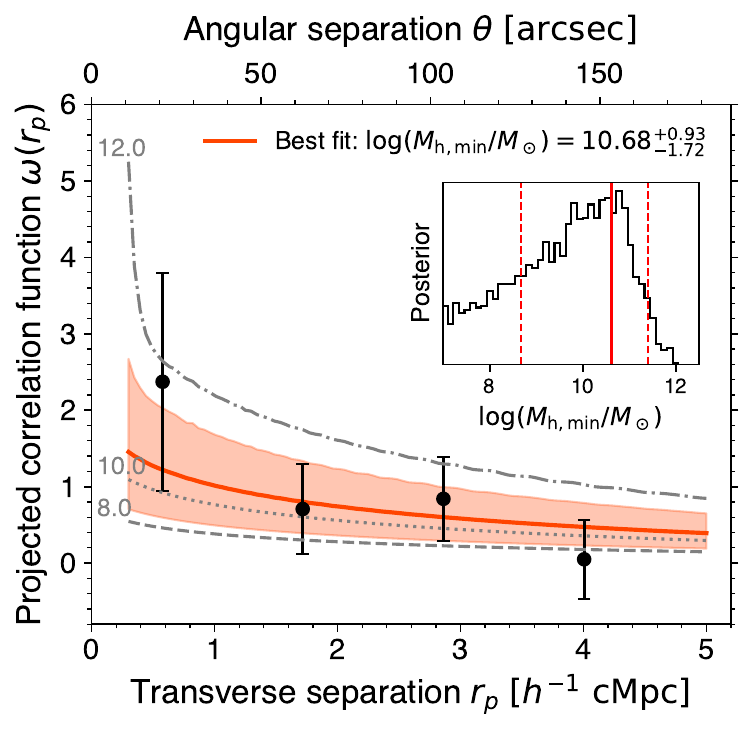}
    \caption{\small The volume-averaged projected Pop III absorber-galaxy cross-correlation function is shown as black data points with $1\sigma$ error bars. The red solid line represents the best-fit HOD model at the mode of the MCMC chains, with the shadow region indicating the $68\%$ Highest Posterior Density Interval (HPDI) of the fitting. The gray lines show models with different $M_{\text{h,min}}$ for comparison. Specifically, $\log(M_{\text{h}}/M_{\odot})=8$ is the threshold for atomic cooling halos\cite{Klessen_23}.
    % Patrick_23,Prole_24
    \label{fig:HOD_fit}}
\end{extendedfigure}

The host halo mass of the Pop III absorber is estimated by comparing the observed cross-correlation function with the Halo Occupation Distribution (HOD) model. The mean numbers of Pop III absorber-host galaxies and \OIII emitters in dark matter halos with a mass of $M_h$ are described by $N_P (M_h)$ and $N_G(M_h)$, respectively. We model the total HOD as a sum of the central and satellite galaxies $N_i(M_h)=N_{{\rm cen},i}(M_h)+N_{{\rm sat},i}(M_h)$. The index $i=\rm P,G$ denotes the Pop III and \OIII emitters, respectively. $N_{{\rm cen},i}(M_h)$ and $N_{{\rm sat},i}(M_h)$ are the mean number densities of central and satellite galaxies. We adopt the 5-parameter HOD model\cite{Zheng_05}, where $N_{\rm cen}(M_h)$ is approximated as a step function:
\begin{equation}
    N_{{\rm cen},i}(M_{\mathrm{h}}) = \frac{1}{2}\left[1 + \mathrm{erf}\left(\frac{\log M_{\mathrm{h}} - \log M_{\mathrm{min},i}}{\sigma_{\log M, i}}\right)\right],
\end{equation}
where $\sigma_{\log M, i}$ reflects the scatter in the luminosity–halo mass relation, and $\log M_{\mathrm{min},i}$ is the minimal halo mass which describes the mass scale at which $50\%$ of halos host a central galaxy.
$N_{{\rm sat},i}(M_h)$ is parameterized as a power law:
\begin{equation}
    N_{{\rm sat},i}(M_h) = N_{{\rm cen},i}(M_h) \left( \frac{M_h}{M_{{\rm sat},i}} \right)^{\alpha_i},
\end{equation}
where $M_{\rm sat}$ is the normalization of the power law and $\alpha_i$ is the power-law index. We model the galaxy-absorber cross-power spectrum $P_{\rm PG}(k)$ using these HODs\cite{Cooray_02} with the halo mass function and bias\cite{Tinker_08,Tinker_10}, and the non-linear matter spectrum\cite{Peacock_96}. The cross-correlation function $\xi_{\rm PG}(r)$ is calculated from the cross-power spectrum via the Fourier transform:
\begin{equation}
    \xi_{\rm PG}(r) = \frac{1}{2\pi^2} \int_0^{\infty} k^2 P_{\rm PG}(k) \frac{\sin kr}{kr} \, dk.
\end{equation}

\subsection{HOD model fitting}

We fit the HOD model to the volume-averaged projected Pop III absorber-galaxy cross-correlation function using the Markov Chain Monte Carlo (MCMC) algorithm. For the \OIII emitters, we assume the minimum halo mass $\log_{10}M_{\rm min,G}/{M_\odot}=10.72$ as measured from the auto-correlation function\cite{Eilers_24}. We fit the minimum halo mass $M_{\rm min,P}$ of the Pop III absorber-host galaxies assuming the Gaussian likelihood defined as $L\propto\exp(-\chi^2)$,
\begin{equation}
    \chi^2 = \sum_{i} \frac{[\bar{\omega}_{\text{obs}}(r_{\text{min},i}, r_{\text{max},i}) - \bar{\omega}_{\text{model}}(r_{\text{min},i}, r_{\text{max},i})]^2}{\sigma^2_{\bar{\omega}}(r_{\text{min},i}, r_{\text{max},i})}
\end{equation}
where $\bar{\omega}_{\text{obs}}(r_{\text{min},i}, r_{\text{max},i})$ and $\bar{\omega}_{\text{model}}(r_{\text{min},i}, r_{\text{max},i})$ are the observed and model values of the volume-averaged projected Pop III absorber-galaxy cross-correlation function in the $i$-th radial bin, respectively, and $\sigma_{\bar{\omega}}(r_{\text{min},i}, r_{\text{max},i})$ is the uncertainty of the observed value. The MCMC sampling is performed by the Python package \textsc{emcee}\cite{2013PASP..125..306F}. We assume a fixed $\sigma_{\log M,G}=\sigma_{\log \rm M,P}=0.2$ and fiducial satellite HOD parameters of $\log_{10}M_{\rm sat,G}/{M_\odot}=12$ with $\alpha_G=\alpha_P=1$.
Since it is difficult to constrain all the HOD parameters of Pop III halos with limited data points, we only consider the 2-halo term $P_\text{PG}^{2h}$ and set $P_\text{PG}^{1h}=0$.
We experimented with the parameters for the satellite HODs. Since $P_\text{PG}^{1h}$ only contributes on small scales $(\lesssim 1~h^{-1},\mathrm{cMpc})$, we find that it affects only the innermost radial bin. The estimated minimum halo mass $M_{{\rm min},P}$ changes by less than 0.2 dex, so our conclusions remain unaffected.

In Extended Data Figure \ref{fig:HOD_fit}, we show the best-fit model in comparison with the atomic cooling halo threshold $\log(M_{\mathrm{h,min}}/M_{\odot})=8$. The minimum halo mass associated with the Pop III absorber is estimated to be $\log(M_{\mathrm{h,min}}/M_{\odot})=10.68^{+0.93}_{-1.72}$. The measurement uncertainties also allow for the possibility of lower-mass halos down to $\log(M_{\mathrm{h}}/M_{\odot})\sim9$. We thus put the lower limit of the observed Pop III-enriched host halo mass as $\log(M_{\mathrm{h}}/M_{\odot})>9$. This lower limit remains $\sim1$ dex above the atomic cooling threshold, thereby constraining the host halos of Pop III absorbers in our observation to be relatively massive atomic cooling halos rather than low-mass molecular cooling halos.

From our best-fit halo mass, the corresponding (physical) virial radius\cite{Mo_10} of the Pop III absorber host halo is:
\begin{equation}
    r_{\text{vir}} \approx 17\,\text{pkpc} \left[\frac{M_{\text{h}}}{10^{10.68}M_{\odot}}\right]^{1/3} \left[\frac{\Delta_{\text{vir}}}{200}\right]^{-1/3} \left(\frac{1+z}{6.945}\right)^{-1},
\end{equation}
where $\Delta_{\text{vir}}$ is the overdensity of dark matter halos. Assuming $\Delta_\text{vir}=200$, and considering the $1\sigma$ uncertainty of the halo mass, we obtain $r_\text{vir}=17^{+18}_{-12}\,\rm pkpc$. Suppose we associate the nearest \OIII emitter with the host halo of the Pop III absorber. In this case, the physical distance from the emitter to the absorber is 238 pkpc (with a projected distance 119 pkpc), suggesting a relic of Pop III star formation at the outskirts ($\sim14^{+34}_{-7}~r_{\rm vir}$) of a massive galaxy, similar to what is predicted by cosmological simulations\cite{Zier_25}. It remains possible that closer, fainter halos exist below the current detection limit and may also contribute to the Pop III formation.

\end{methods}

\begin{addendum}

\item[Data availability] 
The JWST data are available in the Mikulski Archive for Space Telescopes (MAST; \url{http://archive.stsci.edu}), under JWST programs GO-2078\cite{ASPIREdata} and GO-1243\cite{EIGERdata}. The metal absorber catalog of XQR-30 is available at \url{https://github.com/XQR-30/Metal-catalogue}. The \OIII emitter catalog of the EIGER survey is available at \url{https://eiger-jwst.github.io/data.html}. Other data presented in this paper are available upon request to the corresponding author, Zihao Li.

\item 
The authors thank the anonymous referees for their constructive comments, which helped improve the manuscript.
This work is based on observations made with the NASA/ESA/CSA James Webb Space Telescope, associated with the programs GO-1243 and GO-2078, and Very Large Telescope associated with ESO programs: X-shooter 0100.A-0625, 0102.A-0154, 1103.A-0817, and 096.A-0095. We acknowledge the University of Copenhagen for awarding this project access to the LUMI supercomputer, owned by the EuroHPC Joint Undertaking, hosted by CSC (Finland) and the LUMI consortium through Denmark. Z.L. thanks Yi Ren for helpful discussions.
We thank Oliver Zier, Aaron Smit, Rahul Kannan, and Enrico Garaldi for helpful discussions on the \textsc{thesan-zoom} simulation.

\item[Funding Statement]
Z.L. acknowledges the fellowship from the Cosmic Dawn Center, which is funded by the Danish National Research Foundation under grant no. 140. K.K. acknowledges support from VILLUM FONDEN (71574). L.C. is supported by DFF/Independent Research Fund Denmark, grant-ID 2032–00071. V.D. acknowledges ﬁnancial support from the Bando Ricerca Fondamentale INAF 2022 Large Grant “XQR-30”. A.V. acknowledges funding from the Cosmic Frontier Center and the University of Texas at Austin’s College of Natural Sciences. S.Z. acknowledges support from the Chinese Academy of Sciences (no. E5295401).

\item[Author Contributions]
Z.L. reduced the JWST data, conducted the analysis, and wrote the manuscript. K.K. and L.C. conceived the project and contributed to both the analysis and the manuscript preparation. Z.C. contributed to the discussions. V.D. led the XQR-30 program and contributed to the data analysis. J.M, D.K., R.B., and R.M. contributed to the EIGER data/catalog and discussions. F.W. contributed to the ASPIRE data. T.B., F.D., and S.Z. contributed to the ionization correction analysis. I.V. and S.S. contributed to the metal enrichment model and manuscript writing. A.V. contributed to the theoretical interpretations. S.B., E.B., F.D., X.F., H.J., X.J., M.L., S.R., J.Y., S.Z., H.Z. and Y.Z. contributed to the manuscript through comments and discussions.

\item[Competing Interests]
The authors declare no competing interests.
\end{addendum}

\section*{References}
\par

%-----------------------------------------------------------------%
% \clearpage

% \bibliography{reference}{}
% \bibliographystyle{naturemag}

\end{document}